\documentclass[aps,prb,twocolumn,reprint,superscriptaddress,amsmath,amssymb]{revtex4}
\usepackage{bm}
\usepackage{graphicx}
\usepackage{array}
\usepackage[bookmarks=false]{hyperref}

\usepackage{color}

\begin{document}

\title{Exchange splitting and exchange-induced non-reciprocal photonic behavior of graphene in CrI$_3$-graphene vdW heterostructures}

\author{Alexander M. Holmes}
\email[]{holmesam@uwm.edu}
\affiliation{Department of Electrical Engineering, University of Wisconsin-Milwaukee, 3200 N. Cramer St., Milwaukee, Wisconsin 53211, USA}

\author{Samaneh Pakniyat}
\email[]{pakniyat@uwm.edu}
\affiliation{Department of Electrical Engineering, University of Wisconsin-Milwaukee, 3200 N. Cramer St., Milwaukee, Wisconsin 53211, USA}

\author{S. Ali Hassani Gangaraj}
\email[]{ali.gangaraj@gmail.com}
\affiliation{School of Electrical and Computer Engineering, Cornell University, Ithaca, NY 14853, USA}

\author{Francesco Monticone}
\email[]{francesco.monticone@cornell.edu}
\affiliation{School of Electrical and Computer Engineering, Cornell University, Ithaca, NY 14853, USA}

\author{Michael Weinert}
\email[]{weinert@uwm.edu}
\affiliation{Department of Physics, University of Wisconsin-Milwaukee, Milwaukee, Wisconsin, 53211, USA}

\author{George W. Hanson}
\email[]{george@uwm.edu}
\affiliation{Department of Electrical Engineering, University of Wisconsin-Milwaukee, 3200 N. Cramer St., Milwaukee, Wisconsin 53211, USA}

\date{\today}

\begin{abstract}
The exchange splitting and resulting electromagnetic response of graphene in a
monolayer chromium triiodide (CrI$_3$)--graphene van der Waals (vdW) heterostructure
are considered using a combination of density-functional theory and electromagnetic
calculations. Although the effective exchange fields are in the hundreds of Tesla, for the equilibrium separation non-reciprocal effects are found to be weak compared to those for a comparable external magnetic bias. For non-equilibrium separations, nonreciprocal effects can be substantial.

\end{abstract}

\maketitle


\section{Introduction}
Graphite consists of parallel atomic layers of carbon atoms, the layers
being weakly bound together by van der Waals (vdW) forces. As such, graphite
is easily cleaved to form few layer materials, or even monolayers (graphene). Since its experimental isolation in 2004 \cite{Gro}, graphene has
been an object of considerable study for both scientific and industrial
investigators. Graphene's most notable feature is its atomic hexagonal lattice,
which results in linear electronic dispersion and the presence of Dirac
points at the Fermi level. As a result, electrons behave as massless particles in the vicinity
of the Dirac point, leading to extraordinary electrical and mechanical
properties \cite{GR}.

Bulk chromium triiodide, CrI$_3$, also is a layered vdW material that can be easily
cleaved, and is relatively stable in ambient conditions \cite{BSA}. Bulk CrI$_3$
is a ferromagnetic (FM) insulator with a relatively high Curie
temperature of T$_\mathrm{c}$=61 K \cite{BSA}. The 2D/monolayer
form of CrI$_3$ consists of Cr$^{3+}$ ions
and I$^{-}$ ions that form edge-sharing octahedra arranged in a hexagonal
honeycomb lattice with an approximate thickness of 0.6 nm. Like its bulk form, monolayer CrI$_3$ is also a FM insulator, with an
out-of-plane easy axis and somewhat reduced T$\mathrm{c}$ of 45 K \cite{SupTc}. 

The controlled growth/deposition of 2D materials can lead to van der Waals heterostructures that result in exceedingly thin
structures with enhanced functionality. Here, we exploit
the proximity exchange between a 2D ferromagnet and graphene. In its monolayer form, CrI$_3$ exhibits massive local Cr magnetic
moments of 3$\mu _\mathrm{B}$, 
which potentially can induce large exchange splittings in adjacent layers of a
heterostructure. Since 2D CrI$_3$ has a hexagonal structure, it is well
lattice-matched with graphene. Magnetic order in CrI$_3$ has been studied
experimentally in \cite{35,36,39,42,43}, and in other 2D magnets, such as MnSe%
$_{2}$ \cite{27,28} and CrGeTe$_{3}$ \cite{29,30}. In all
cases, these 2D magnets have out of the plane magnetization. In some cases,
magnetic effects can be controlled via electrostatic gating \cite{42,43}, or strain \cite{PE,Chern1}. 

Enormous
pseudo-magnetic fields (on the order of hundreds of Tesla) and associated
pseudo-Landau levels (LLs) have been predicted in strained systems \cite{PMF}. Such fields do not break time-reversal (TR)
symmetry, and can not lead to nonreciprocal behavior. Importantly, the exchanged-induced fields described here do break TR: The effective Hamiltonians for  both an external magnetic
field and a ferro-/antiferro-magnetic system contain terms that
explicitly couple to the spin that are not invariant under time
reversal; in contrast, the pseudo-magnetic fields in strained graphene
couple to charge only, and hence preserve time-reversal symmetry. Exchange interactions in similar vdW heterostructures have been
considered, e.g., Cr$_{2}$Ge$_{2}$Te$_{6}$-graphene \cite{crgr1}, where
equilibrium exchange splittings were calculated to be approximately 5 meV, and EuS-graphene \cite{SEF}. A Chern insulating state can be
realized in graphene in proximity to CrI$_3$, via the magnetic exchange field
and Rashba spin-orbit coupling (SOC) \cite{CIS,Chern1}. However, to
achieve this, the heterostructure needs to be compressed from its
equilibrium state which increases the effective field \cite{CIS}.

In this work, we use first-principles density functional theory (DFT)
calculations to show that the proximity exchange in graphene due to
monolayer CrI$_3$ can result in an enormous exchange field, and then we investigate
the conductivity of graphene due to the CrI$_3$ exchange field, and the behavior of
bulk and nonreciprocal edge surface-plasmon polaritons (SPPs). A comparison is made
with the conductivity and SPP properties of graphene in an external magnetic field,
and significant differences are found in the two cases. We also examine Faraday
rotation (FR) of graphene \cite{FRG}-\cite{SIFR}. The principal findings of this
work are: (1) the equilibrium (minimum energy) separation between the CrI$_3$ and
graphene is approximately 3.75 \AA, at which point the exchange splitting is 21 meV,
corresponding to an effective exchange field of 100 T and a chemical potential of
$\mu=-0.3$eV, which self-biases the graphene. Referring to graphene's conductivity
in the CrI$_3$-graphene heterostructure, (2) Landau levels, which are the most
prominent feature of the graphene conductivity in a strong external field, are
absent in the case of the exchange field. (3) In the far-infrared considered here,
the intraband conductivity is dominant, with diagonal element values that are
approximately the same as isolated graphene with no applied magnetic bias and
$\mu=-0.3$eV, whereas the off-diagonal elements are similar in magnitude to those in
the external bias case. (4) Because of the large diagonal conductivity response compared to having an external bias (in which case most of the Drude weight is transferred to the Landau levels), the resulting non-reciprocity due to the exchange field is considerably less then for an external magnetic field of the same strength. For smaller separation (achievable through, e.g., strain), nonreciprocal effects in Faraday rotation are still rather modest, but a unidirectional edge SPP can be found.

The article is organized as follows. In Section \ref{DFT} the density functional
calculations are presented, and results for exchange splittings and the
corresponding effective exchange fields are given. In Section \ref{ExcSigma} the
exchange-field-induced graphene conductivity is discussed, and compared with that
arising from an external bias, and bulk and edge surface plasmons are considered.
The edge SPPs for the exchange field are slightly non-reciprocal for the equilibrium separation, whereas for the external bias case they are highly nonreciprocal (unidirectional), tightly-confined, long-lasting, and robust to material discontinuities. In Section \ref{FRS}, Faraday
rotation is shown for both the exchange and external bias fields, where, again, the
exchange field is shown to produce modest Faraday rotation. The Supplemental
Information contains further results from the DFT calculations, and the
derivation of the edge plasmon dispersion. In the following, the suppressed time dependence is $e^{-i\omega t}$. 

\section{Density Functional Calculations}\label{DFT}

\begin{figure}
  \includegraphics[width=1.0\columnwidth]{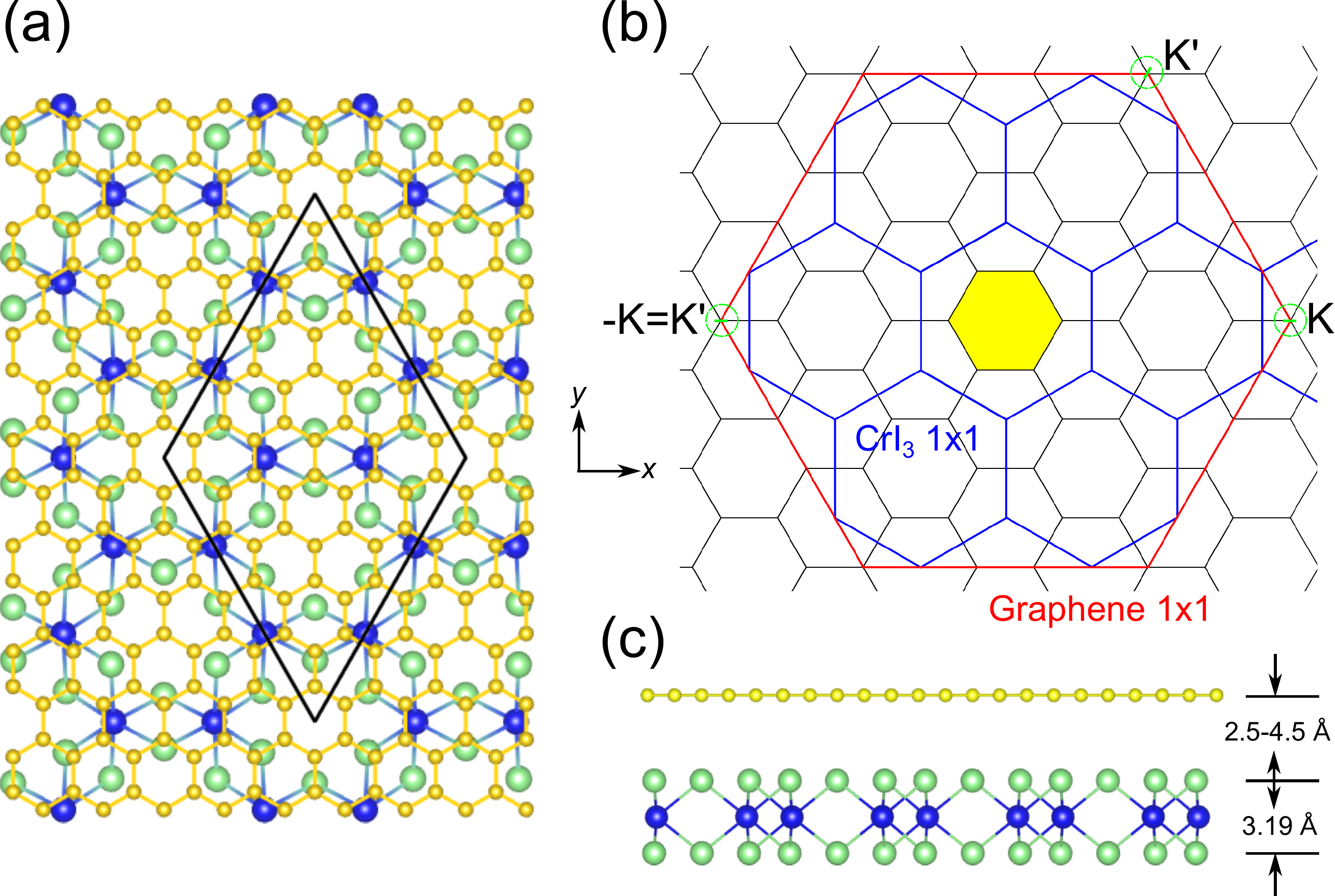}
  \caption{
(a) Top and (c) side views of the ($\sqrt{3}$$\times$$\sqrt{3}$) CrI$_3$--(5$\times$5)
graphene structure (C: yellow; Cr: blue; I: green), with the commensurate
supercell given in black.
(b) The Brillouin zones of the supercell (black; first Brillouin zone in
yellow), (1$\times$1) CrI$_3$ (blue), and (1$\times$1) graphene (red). The green
circles (and enclosed lines) denote the portion of $k$-space where the graphene
Dirac points occur.
  }
 \label{fig-struct}
\end{figure}

Density functional calculations for graphene on CrI$_3$ have been reported
previously, \cite{Chern1}, with an emphasis on the topological aspects of the
compressed system. Here the focus is on the effect of the induced exchange field
on the graphene electronic structure, and the implications for the calculation
of optical properties discussed later; detailed first-principles calculations
of the optical properties of the CrI$_3$ itself have also been
reported previously \cite{CrI3-Louie}.

To model the composite system, we consider 5$\times$5 graphene on a
free-standing $\sqrt{3}$$\times$$\sqrt{3}$ CrI$_3$ monolayer,
Fig.~\ref{fig-struct}(a), which has a small lattice mismatch of $\sim$1\%.
Because graphene and CrI$_3$ are both layered van der Waals materials, 
the results presented below are only very weakly dependent on the horizontal
registry between the two, as verified by considering two different less symmetric
registries. The
supercell Brillouin zone, which is a factor of 25 (3) times smaller than that
of graphene (CrI$_3$), is shown in Fig.~\ref{fig-struct}(b).
The interlayer separation is varied between 2.5 and 4.5 \AA. 

\begin{figure}
  \includegraphics[width=1.0\columnwidth]{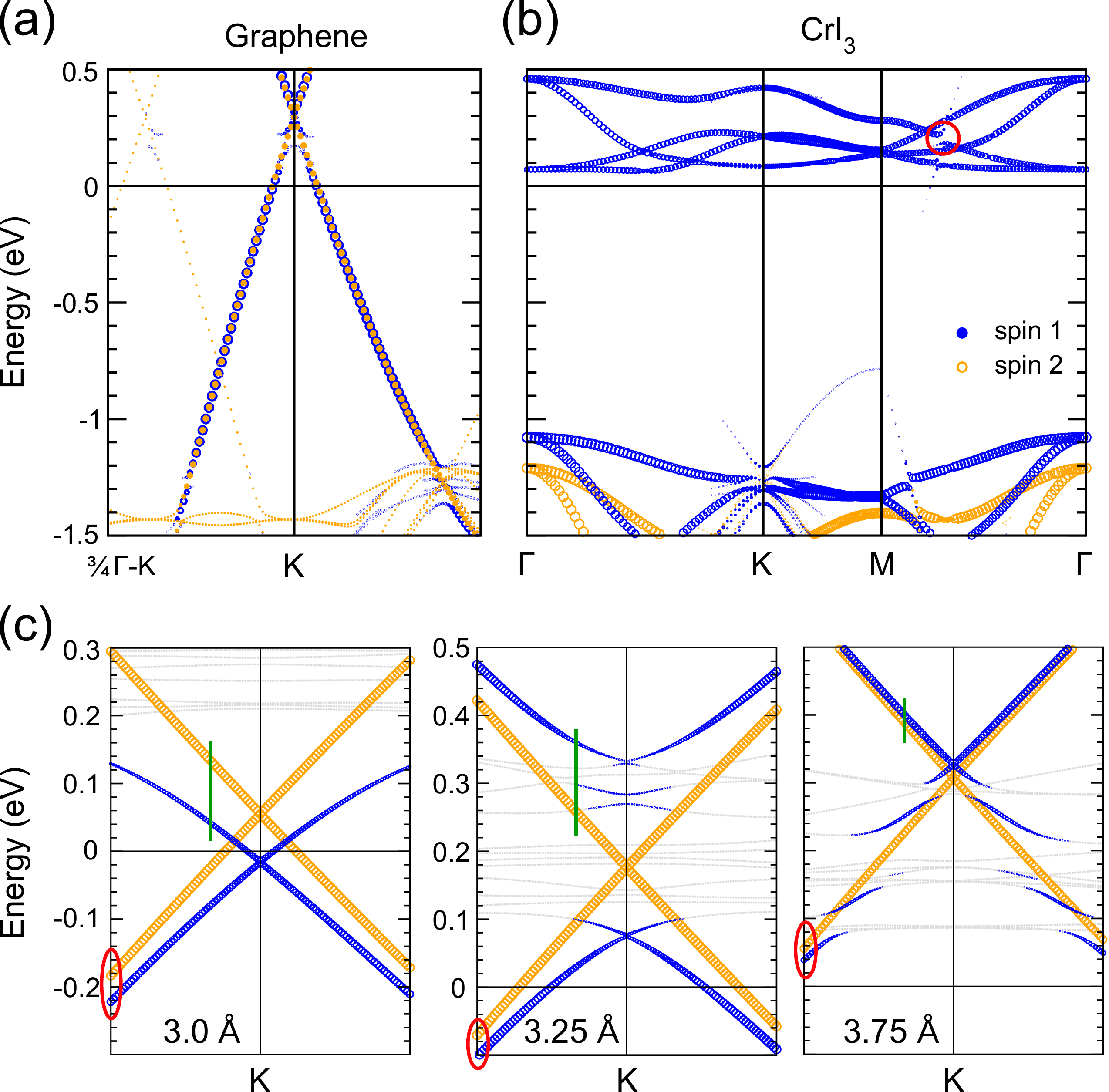}
  \caption{$k$-projected bands of the graphene--CrI$_3$ magnetic system.
Blue (orange) denote the majority (minority) spin of the CrI$_3$, and the size
of the circles represents the relative weight.
(a) Graphene $k$-projected bands around the K point
($\pm\frac{1}{4}$) along $\Gamma$-K of the (1$\times$1)
Brillouin zone, and
(b) CrI$_3$ $k$-projected bands along the high symmetry directions of the
(1$\times$1) structure, both for a graphene--CrI$_3$ separation of 3.75\,\AA.
(c) Close-ups of the graphene $k$-projected bands bands within $\pm\frac{1}{40}$
of K for different separations. The gray bands are (``folded'' and CrI$_3$)
bands with small weights. The green lines and red ovals show where the
exchange splittings above and below the Dirac point, respectively, are measured.
  }
 \label{fig-bands}
\end{figure}

The calculations were performed using the Vienna Ab initio Simulation Package
(VASP) \cite{vasp} within the GGA+\textit{U} rotationally invariant
approximation \cite{LDAU}, with the choice of $U$ = 1 eV, $J$=0.  (As shown in
Ref.~[\onlinecite{CrI3-Louie}], the CrI$_3$ gap decreases with increasing $U$,
contrary to normal expectations. This choice of parameters provides a reasonable
starting point for the CrI$_3$ electronic structure.)
In addition, van der Waals DFT-D3 corrections \cite{DFT-D3} were included. The
projector-augmented wave functions were expanded in plane waves up to 400 eV,
the repeated slab geometry included a vacuum region of at least 20 \AA, and a
27$\times$27$\times$1 $k$-point mesh in the supercell was used for
self-consistency, corresponding to a 135$\times$135$\times$1 mesh for
graphene; this mesh was sufficient for placing the Fermi level and for
the optical conductivity calculations.
Spin-orbit was included for some
calculations (see Supplemental Information) using the same parameters.

Each ferromagnetically coupled Cr has a magnetic moment of 3 $\mu_\mathrm{B}$, 
and the moments are calculated ($\sim$0.4 meV/Cr) to be orientated perpendicular
to the plane. The spin-polarized \textit{k}-projected \cite{kproj1,kproj2} bands of graphene around the
K point and of CrI$_3$ at the calculated equilibrium graphene-CrI$_3$ separation
of 3.75 \AA{} are shown in Figs.~\ref{fig-bands}(a) and (b), respectively. The
top of the CrI$_3$ valence band and the lowest set of conduction bands are of
majority spin (blue curves). The
graphene Dirac point lies above the Fermi level in the conduction band of
CrI$_3$, and opens up a gap in the CrI$_3$ bands along $\Gamma$-M (red circle in
Fig.~\ref{fig-bands}(b)). The relative position of the graphene and CrI$_3$
bands varies with interlayer separation (Fig.~\ref{fig-bands}(c) and
Supplemental Information Fig.~S2): for separations of less than $\sim$3.2 \AA,
the graphene Dirac point is in the gap and then crosses into the CrI$_3$
conduction band. This behavior can be understood by noting that the calculated work
functions and CrI$_3$ gap place the graphene Dirac point within the CrI$_3$
conduction band, for both the present GGA+\textit{U}
calculations and for hybrid HSE functional calculations \cite{Chern1}.
However, GW calculations for CrI$_3$ \cite{CrI3-Louie} increase the size of the
gap, so that at the equilibrium separation the Dirac point may still be within
the gap. Regardless, the present results can provide insight into the difference
in expected behavior with the relative placement of the two sets of bands.

For all separations, Fig.~\ref{fig-bands}(c), the minority (``spin 2'') graphene
bands maintain their linear dispersions, even including spin-orbit interaction
(c.f., SI Figs.~S3 and S4). The majority bands, on the other hand, interact and
hybridize with the (majority spin) conduction band states even for smaller
separations where the Dirac point is in the CrI$_3$ gap.
Importantly, because of the proximity of the
graphene to the ferromagnetic CrI$_3$, there are induced exchange splittings of
the graphene bands. 
For larger separations,
the majority graphene bands that overlap the GrI$_3$ conduction bands are strongly
modified, whereas the minority bands retain the characteristic graphene dispersions.

\begin{figure}
  \includegraphics[width=1.0\columnwidth]{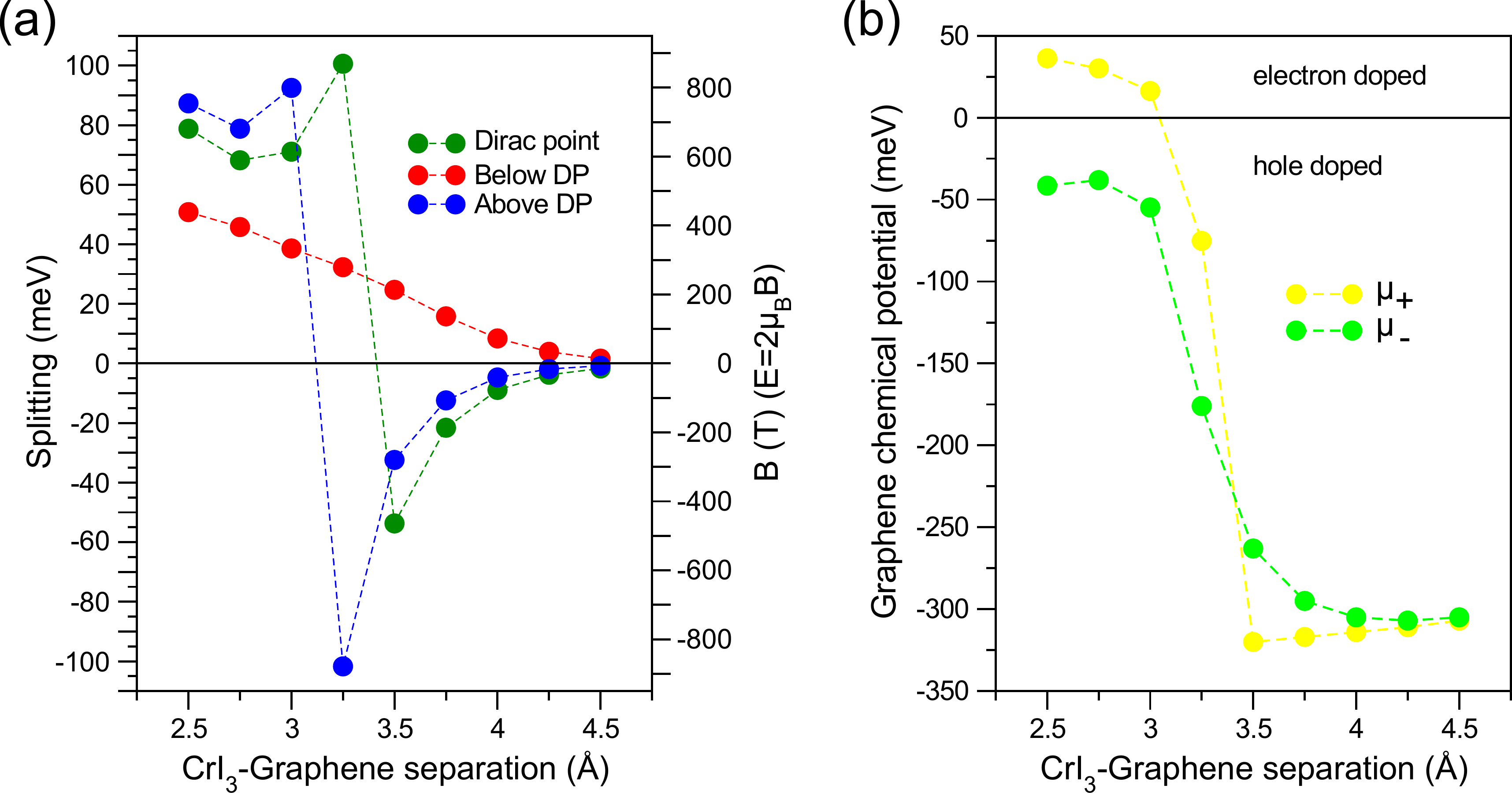}
  \caption{
(a) Exchange splittings (meV) and effective field (T), of the graphene states
around K for different separations determined at the positions shown in
Fig.~\ref{fig-bands}(c) (above and below the Dirac point indicates the green line and red ovals, respectively, in Fig.~\ref{fig-bands}(c)). (b) The chemical potentials of graphene for each spin relative to the respective
Dirac points.
  }
 \label{fig-mu}
\end{figure}

The calculated splittings of the Dirac point and the bands above (below)
measured at the indicated positions are given in Fig.~\ref{fig-mu}(a). These
splittings are large compared to the Zeeman splittings induced by an external
field: the effective fields are in the range of 100 T. When the Dirac point is
in the gap, the exchange splittings are normal in the sense that the majority
states are deeper in energy than the minority. However, the exchange splitting
of the Dirac point and the bands above reverse as the Fermi level of the
combined system moves into the CrI$_3$ conduction band.

Because of the exchange splitting and the relative positions of the bands, the
graphene is effectively doped, which can be described by spin-dependent chemical
potentials, $\mu_\pm$, as shown in Fig.~\ref{fig-mu}. For smaller separations
with the Dirac points in the gap, $\mu_\pm$ are approximately equal and of
opposite sign, i.e., no net doping. For larger separations, including the
equilibrium one, the graphene becomes hole doped with  $\mu_\pm$$\sim$0.3 eV.

For graphene in external magnetic fields and non-zero chemical potential,
the intraband contributions to $\sigma_{xx}$ dominate over interband ones in the
far-infrared optical conductivity, and
the formation of Landau levels provide an explanation of the Hall
conductivity $\sigma_{xy}$.
Although the \textit{effective} fields due to the proximity-induced
exchange splittings are large, these do not create Landau levels; the
formations of the minibands in the majority bands seen in
Fig.~\ref{fig-bands}(c) are due to interactions and hybridization with
the CrI$_3$. The Landau levels formed in graphene in the presence
of external magnetic fields or strain-induced pseudomagnetic fields \cite{sciadv2019} are both more localized in energy and have their broad momentum
distribution peaked around K. Similar to Landau levels, however,
these minibands change the dispersion and hence will modify the optical transitions.

\section{Graphene Conductivity and Surface and Edge Plasmon Polaritons Due to CrI$_3$ Exchange Field}\label{ExcSigma}

\begin{figure}[b]
		\includegraphics[width=0.95\columnwidth]{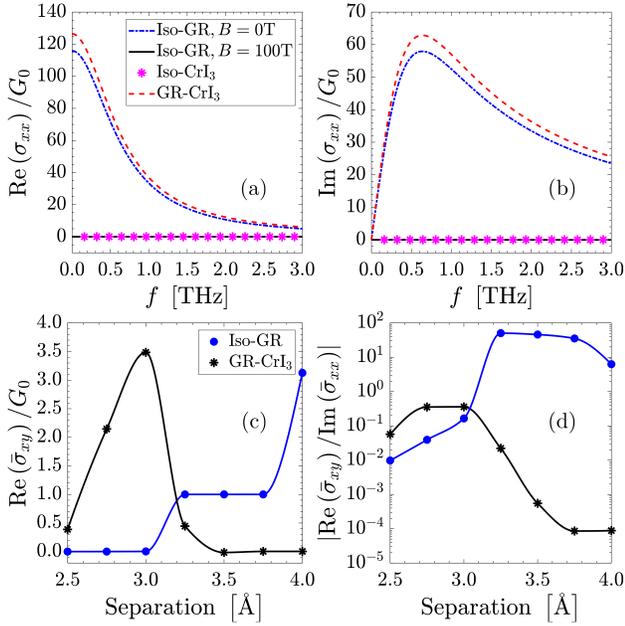}
		\caption{Panels a, b: Two-dimensional conductivity of graphene in the presence of the CrI$_3$ exchange field at the equilibrium separation of 3.75\AA.  Also shown for comparison is the conductivity of isolated graphene (Iso-GR) computed assuming an external field of 100T \cite{Gus1} and the zero external field case (in both of the latter computations, $\Gamma=2 \times 10^{12}$/s, $\mu=-0.3$eV, and $T=40$K). Panel c shows the off--diagonal element vs. CrI$_3$-graphene separation compared with the isolated-graphene external bias case (where for each `separation' the value of effective field from Fig. \ref{fig-mu} was assumed), and Panel d shows the off-diagonal element normalized by the frequency average of the diagonal element. In Panels c,d, the over-bar indicates frequency average, as explain in the text. $G_0=2e^2/h$.} 
		\label{GCE}
\end{figure}

The computation of the intra- and inter-band contributions to graphene's
conductivity in the presence of the CrI$_3$ exchange field is briefly described in
the Supplemental Information. Figure \ref{GCE} shows the computed conductivity in
the far-infrared, and, for comparison, the conductivity computed (\textit{i})
assuming isolated graphene in an external bias of 100 T and $\mu=-0.3$eV
\cite{Gus1}, the
effective equilibrium exchange field and potential as described above; 
(\textit{ii})
isolated graphene $B=0$ T, $\mu=-0.3$ eV; and (\textit{iii}) the calculated CrI$_3$ conductivity 
\cite{CrI3-Louie}. The latter is negligible at the considered frequencies. Figures
\ref{GCE}(a, b) show the diagonal elements vs.\ frequency, and Fig.~\ref{GCE}(c) shows the off-diagonal
element vs.\ CrI$_3$-graphene separation (which are nondispersive in our
calculation). In Fig.~\ref{GCE}(c) for the external bias case,
for each ``separation'' the value of effective field from Fig.~\ref{fig-mu} was
assumed (using the below the DP points).  In Fig.~\ref{GCE}(c,d), we show the mean of conductivity over the 0 to 5THz range. As an example, for the CrI$_3$-graphene data (separation 3.75A), this mean is $33G_0$, and for isolated biased graphene (B0 = 100T), $0.03488G_0$. 

The diagonal elements of the conductivity are dominated by the Drude intraband
contribution at the considered frequencies
($\sigma_{xx}^\mathrm{2D-Drude}={i\Omega}/({\omega+i\Gamma})$, with the Drude
weights $\Omega$ shown in Fig.~S5), resembles very closely the conductivity for
isolated graphene with no magnetic bias, but with the exchange-field induced value
of chemical potential $\mu=-0.3$eV. Because of transfer of the Drude weight to the
Landau levels in the case of an external bias, the exchange field diagonal
conductivity is several orders of magnitude larger than the equivalent external
field conductivity. 

For some separations, c.f., Fig.~\ref{fig-mu}, the chemical potential is quite
different for the two spins, as is the effective bias field. Therefore, for the
external bias computation, we adjust the spins accordingly and sum over the two
spins. For the off-diagonal elements, the exchange field values are similar in
magnitude as the external bias case having the same effective field. Figure
\ref{GCE}d shows the off-diagonal values of conductivity normalized by the frequency
average of the diagonal element, since this ratio is an indication of the
non-reciprocity of the material. Notably, the non-reciprocity of the exchange field
case is much weaker than for the external field bias. 

From Fig.~\ref{GCE}, the CrI$_3$ conductivity is much smaller in magnitude than the
graphene conductivity. Since these effectively combine in parallel, from an
electromagnetic standpoint, we can ignore the presence of the CrI$_3$ in the electromagnetic calculations. This was confirmed by computing the dispersion of the CrI$_3$-graphene system including both conductivities, as in \cite{Hanson2}.

\subsection{Bulk (Surface) SPPs}

For the case of graphene having an arbitrary conductivity tensor and residing in a
homogeneous medium characterized by $\mu_0$ and $\varepsilon$, SPPs of an infinite
2D sheet satisfy a dispersion equation of the form $D\left( k_{x},k_{y}\right) =0$ \cite{Hanson1}, where
\begin{align}
& D\left( k_{x},k_{y}\right) =k_{x}k_{y}\left( \sigma _{yx}+\sigma
_{xy}\right) +\left( k_{y}^{2}-k_{1}^{2}\right) \sigma _{yy}  \label{de} \\
& \ +\left( k_{x}^{2}-k_{1}^{2}\right) \sigma _{xx}-2ip\varepsilon \omega \left(
1+\frac{1}{4}\eta ^{2}\left( \sigma _{xx}\sigma _{yy}-\sigma _{xy}\sigma
_{yx}\right) \right) , \nonumber  
\end{align}%
where $p=\sqrt{k^{2}-k_{1}^{2}}$, $k=\left\vert \mathbf{k}\right\vert $ is the in-plane wavenumber, and $k_{1}=\omega \sqrt{\mu_0 \varepsilon }$. The square root in $p$ leads to a
two-sheeted Riemann surface in the $q$-plane, and associated branch cuts. The
standard hyperbolic branch cuts \cite{Ishimaru} separate the proper (where $\textrm{Re}\left( p\right) >0$, such that the radiation condition as 
$\left\vert z\right\vert \rightarrow \infty$ is satisfied) and improper
sheets.

In the presence of the exchange field, or an external magnetostatic bias, the conductivity tensor elements are 
\begin{align}
\sigma _{xx}& =\phantom{-}\sigma _{yy}=\sigma _{d},  \label{HEC} \\
\sigma _{xy}& =-\sigma _{yx}=\sigma _{o}.  \nonumber
\end{align}%
In this case, (\ref{de}) reduces to%
\begin{equation}
D\left( k\right) =\left( p^{2}-k_{1}^{2}\right) s_{d}-ipk_{1}\left(
1+s_{d}^{2}+s_{o}^{2}\right),  \label{D}
\end{equation}%
where $s_{d}=\eta \sigma _{d}/2$,$\ s_{o}=\eta \sigma _{o}/2$, and$\ \eta =%
\sqrt{\mu_0 /\varepsilon }$, and $D\left(
k\right) =0$ can be solved to yield%
\begin{equation}
k^{\pm }=k_{1}\left[ \frac{1}{2^{2}s_{d}^{2}}\left( -is^{2}\pm \sqrt{%
4s_{d}^{2}-s^{4}}\right) ^{2}+1\right] ^{1/2},  \label{swpc}
\end{equation}%
where $s^{2}=s_{d}^{2}+s_{o}^{2}+1$. For $s_{o}=0$, and Eq.~(\ref{swpc})
becomes
\begin{equation}
k^{+}=k^{TM}=k_{1}\sqrt{1-\frac{1}{s_{d}^{2}}},\ \ k^{-}=k^{TE}=\ k_{1}\sqrt{%
1-s_{d}^{2}},  \label{pcsc}
\end{equation}%
for the transverse-magnetic (TM) and transverse-electric
(TE) cases, respectively, where transverse is defined with respect to the radial
coordinate. For isolated and unbiased graphene characterized by isotropic complex surface
conductivity, $\sigma =\sigma ^{\prime }+i\sigma ^{\prime \prime }$, a proper
TE surface wave exists if and only if $\sigma ^{\prime \prime }<0$, and a
proper TM surface wave exists for $\sigma ^{\prime \prime }>0$. If $\mu \neq 0$,
pure TM and TE modes do not exist in the presence of a magnetic bias, although usually the modes retain similar characteristics (quasi-TM/TE).

\subsection{Edge Surface Plasmon Polaritons}

\begin{figure}[b]
		\includegraphics[width=1.0\columnwidth]{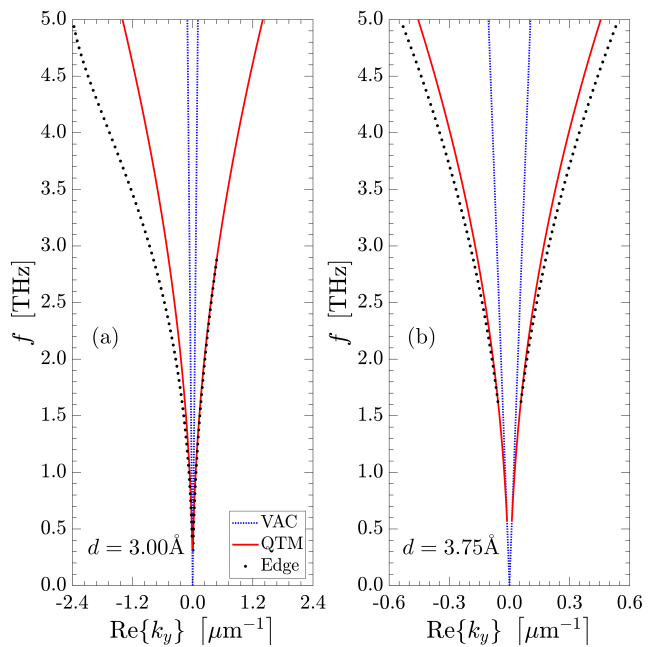}
		\caption{Bulk (solid red) and edge (dots) TM-like dispersion of graphene modes in an exchange field for two CrI$_3$-graphene separations. VAC indicates vacuum dispersion.}  
		\label{DispE}
\end{figure}

Due to symmetry assuming an out-of-plane magnetic bias, bulk SPPs have reciprocal
dispersion; breaking symmetry, for example, by introducing an interface, allows for
plasmons with asymmetrical dispersion \cite{C87}. Specifically, here, we introduce a
graphene half-space, and consider the quasi-static modes that may propagate on the
edge. This is a well-studied problem \cite{Fetter_1}-\cite{Apell}, and here we
consider the exchange field case, and, for comparison, the external bias case. Two
different methods of analysis are presented in the Supplemental Information,
and a new approximate solution for SPP edge dispersion is given there
(Eq.~({56})). 

Figure \ref{DispE} shows the bulk and edge dispersions for a graphene half-space due
to the exchange fields corresponding to separations of 3.0 \AA, Fig.~\ref{DispE}(a),
and 3.75 \AA, Fig.~\ref{DispE}(b). For 3\AA\ separation, the right-going edge mode exists until approximately 3THz, above which the edge mode leaks into the bulk SPP (mathematically, it crosses onto an improper Riemann sheet through a branch point associated with the bulk mode wavenumber); the leaky mode (not shown) then approximately follows the bulk dispersion, with slightly lower wavenumber. In this case, the edge mode is strongly nonreciprocal (unidirectional). However, for the equilibrium separation of 3.75\AA\ separation, the edge mode is essentially reciprocal. 

\begin{figure}[t]
		\includegraphics[width=0.9\columnwidth]{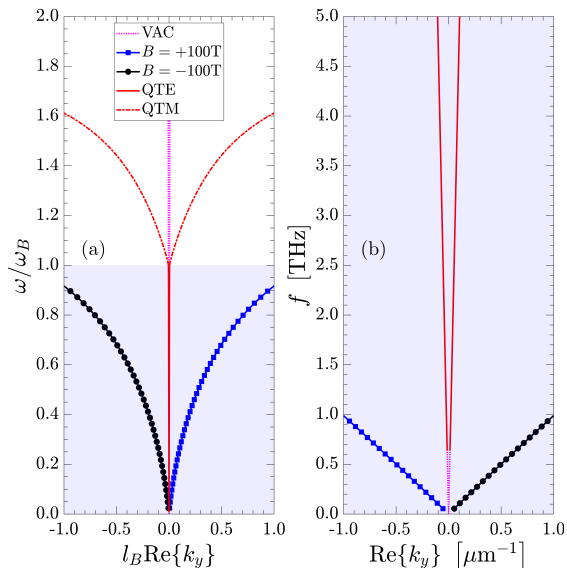}
		\caption{Bulk (dashed red) and edge (dots) TM-like dispersion and bulk TE-like modes (solid red) for graphene in an external bias. The shaded region indicates the bulk bandgap, and $\omega_B=526.2\times 10^{12}$ is the frequency of the first Landau level. $\mu=-0.3$eV, $T=40$K, $\Gamma=2 \times 10^{12}$/s, $B=100$T.  $l_{B}=\sqrt{\hslash /eB}=2.6$nm is the magnetic length.}  
		\label{Disp1}
\end{figure}

The bulk and edge dispersions for graphene in an external magnetic bias field are shown in
Fig.~\ref{Disp1}. The edge modes flip directions upon reversing the bias field. Although the results were computed assuming $B$=100 T, due to the normalization the dispersion diagrams are essentially independent of $B$ for $\left\vert B \right\vert \gtrapprox 1$ T.  For the external bias case, Landau levels are given by 
\begin{equation}
M_{n}=\sqrt{2nv_{F}^{2}\left\vert eB\right\vert \hslash }\approx 36.3\textrm{meV}\sqrt{n\left\vert B\right\vert  },
\end{equation}%
where $-e$ is the charge of an electron, $\hslash =h/2\pi $ is the
reduced Planck's constant, and $v_{F}\simeq 10^{6}$ m/s is the electron Fermi velocity. 

For the exchange field (Fig.~\ref{DispE}), the bulk SPPs are not gapped, whereas for the external bias case
(Fig.~\ref{Disp1}) the bulk SPPs are strongly gapped. This is a result of the behavior of $\textrm{Im}(\sigma)$; Since TM and quasi-TM modes require $\textrm{Im}(\sigma)>0$ for a proper surface wave, gaps appear for $\textrm{Im}(\sigma)<0$, which does not occur for the exchange case in the far-infrared, where the conductivity dispersion is Drude-like. In the external bias case, the formation of Landau levels causes this sign change at lower frequencies, resulting in the TM gap shown in Fig.~\ref{Disp1}.   

\begin{figure}[t]
	\includegraphics[width = 1\columnwidth]{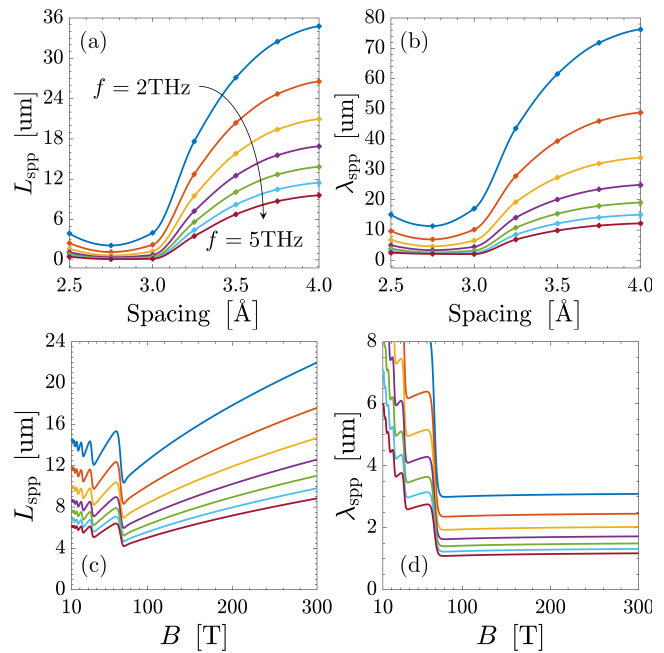}
		\caption{(a) Edge SPP propagation length (left branch of Fig. \ref{DispE}) and (b) SPP wavelength on graphene in an exchange field as CrI$_3$-graphene separation varies. (c) Edge SPP propagation length and (d) SPP wavelength on graphene in an external bias for different frequencies. $\mu=0.05$eV, $T=40$K, $\Gamma=2 \times 10^{12}$/s. For (c)-(d), the first LL occurs at $8.8\sqrt{\left \vert B \right \vert}$, well beyond the considered frequencies. For comparison, from Fig. \ref{fig-mu}, for the separations 2.5\AA, 3\AA, 3.5\AA, 4\AA, the effective bias fields are 750T, 600T, 450T, and 90T, respectively.}  
		\label{LSPP}
\end{figure}

Figure \ref{LSPP}(a,b) shows edge SPP propagation length and guided wavelength on the
graphene layer as a function of CrI$_3$-graphene separation. The SPP propagation
length $1/2\, \textrm{Im}(k_y)$  generally increases with separation, and decreases
with increasing frequency. The SPP wavelength is quite long, ${L}_\textrm{SPP}/\lambda_\textrm{SPP}\ll 1$, and so the SPP seems to be not very useful.

The corresponding edge SPP propagation length and wavelength on the graphene layer in an external
field as a function of the external bias are shown in Fig.~\ref{LSPP}(c,d).
As magnetic bias increases, the SPP propagation length increases, and
${L}_\textrm{SPP}/\lambda_\textrm{SPP}> 1$; for large magnetic bias,
$L_\textrm{SPP}/ \lambda_{\textrm{SPP}} \approx 6-8$. For $\mu=-0.3$eV, the results
are the same as shown in Fig.~\ref{LSPP} for $B>80$ T, since for larger chemical
potentials the SPP is not well-formed and is not quasi-TM below a critical bias \cite{CLB}.

\begin{figure}[t]
		\includegraphics[width=0.95\columnwidth]{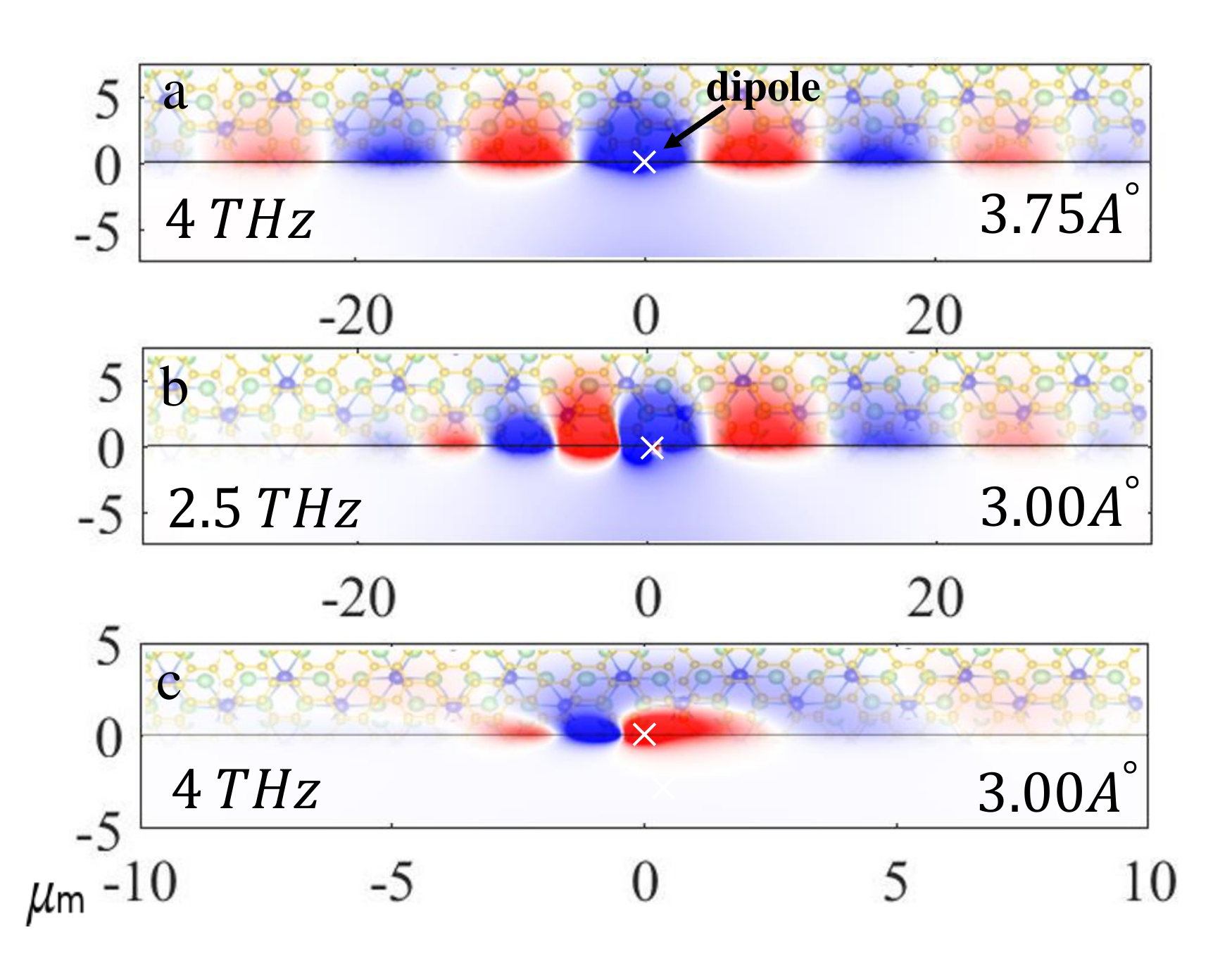}
		\caption{Edge SPP launched by a dipole source near the graphene-vacuum edge for the exchange field magnetic bias for two values of CrI$_3$-graphene separation. }
		\label{discon_exc}
\end{figure}

Figure \ref{discon_exc} shows the edge SPP on the exchange-field biased graphene due
to a dipole source in the vicinity of the graphene-vacuum edge, computed using
COMSOL. In correspondence with the dispersion shown in Fig. \ref{DispE}, for the equilibrium separation of 3.75\AA\ the SPP is essentially reciprocal, as it is at 2.5 THz for separation 3\AA. However, for 3\AA\ and 4 THz, the SPP is unidirectional. However, because $L_\textrm{SPP}/ \lambda_{\textrm{SPP}}$ is short, the SPP does not propagate well. 

\begin{figure}[b]
		\includegraphics[width=1\columnwidth]{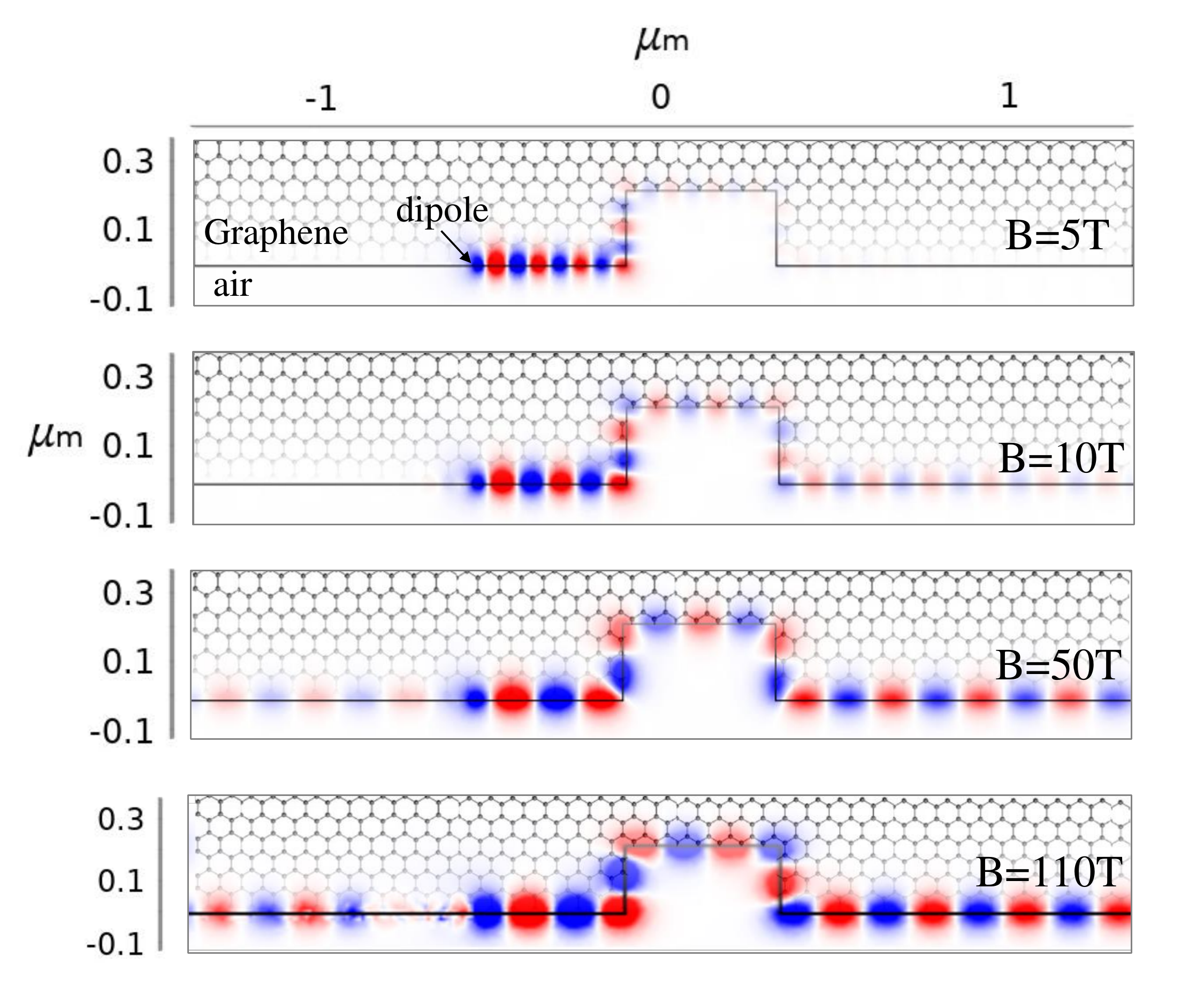}
		\caption{Edge SPP launched by a dipole source near the graphene edge for several values of external magnetic bias. $f=13.87$ THz, $\mu=0.05$eV, $T=40$K, $\Gamma=2 \times 10^{12}$/s.}
		\label{discon}
\end{figure}

In contrast to Fig.~\ref{discon_exc}, Fig.~\ref{discon} shows the edge SPP on
externally-biased graphene due to a dipole source in the vicinity of the edge,
computed using COMSOL. The size of the discontinuity is on the order of
$\lambda_\textrm{SPP}$ (e.g., the length of the discontinuity contour in the second
panel is $5\lambda_\textrm{SPP}$). It is clear that as magnetic bias increases, the
SPP propagates further, in agreement with Fig.~\ref{LSPP}, while its wavelength
increases. The edge SPP is clearly robust, and propagates around the discontinuity.
Although there appears to be a weak field to the left of the source, it is due to
the imperfect boundary condition at the edge of the computational domain.
(Converting to the time domain shows that the field to the left of the source is
actually traveling towards the right.)

\section{Faraday Rotation} \label{FRS}

Faraday rotation on magnetically-biased graphene has been studied in \cite{FRG,GFR},
among other works. As discussed above, since the conductivity of the CrI$_3$ is much
smaller than the conductivity of graphene at the considered frequencies, we can
neglect the Faraday rotation contribution of the CrI$_3$, and only consider the FR
of graphene under the influence of the exchange field. The Faraday rotation is computed as \cite{FR1}
\begin{equation}
\theta_{\textrm{FR}}=\frac{1}{2}\arg \left(\frac{t_{pp}-it_{ps}}{t_{pp}+it_{ps}}\right),
\end{equation}
and the ellipticity is
\begin{equation}
\delta=\frac{\left\vert t_{pp}-it_{ps}\right\vert ^{2}-\left\vert
t_{pp}+it_{ps}\right\vert ^{2}}{\left\vert t_{pp}-it_{ps}\right\vert
^{2}+\left\vert t_{pp}+it_{ps}\right\vert ^{2}},
\end{equation}
where $t_{pp}={E_{p}^{t}}/{E_{p}^{i}}$ and $t_{ps}={E_{s}^{t}}/{E_{p}^{i}}$,
$p=x,y$, $s=x,y$, $s \neq p$, and the superscripts indicate incident ($i$) or
transmitted ($t$) fields. For graphene in a homogeneous medium, the transmission coefficients are
\begin{align}
t_{xx}& =\frac{4/\eta +2\sigma _{xx}}{4\sigma _{xx}+\eta \left( \sigma
_{xx}^{2}+\sigma _{xy}^{2}\right) +4/\eta }, \\
t_{xy}& =\frac{2\sigma _{xy}}{4\sigma _{xx}+\eta \left( \sigma
_{xx}^{2}+\sigma _{xy}^{2}\right) +4/\eta },
\end{align}%
where $\eta =\sqrt{\mu /\varepsilon }$.

\begin{figure}[t]
		\includegraphics[width=0.90\columnwidth]{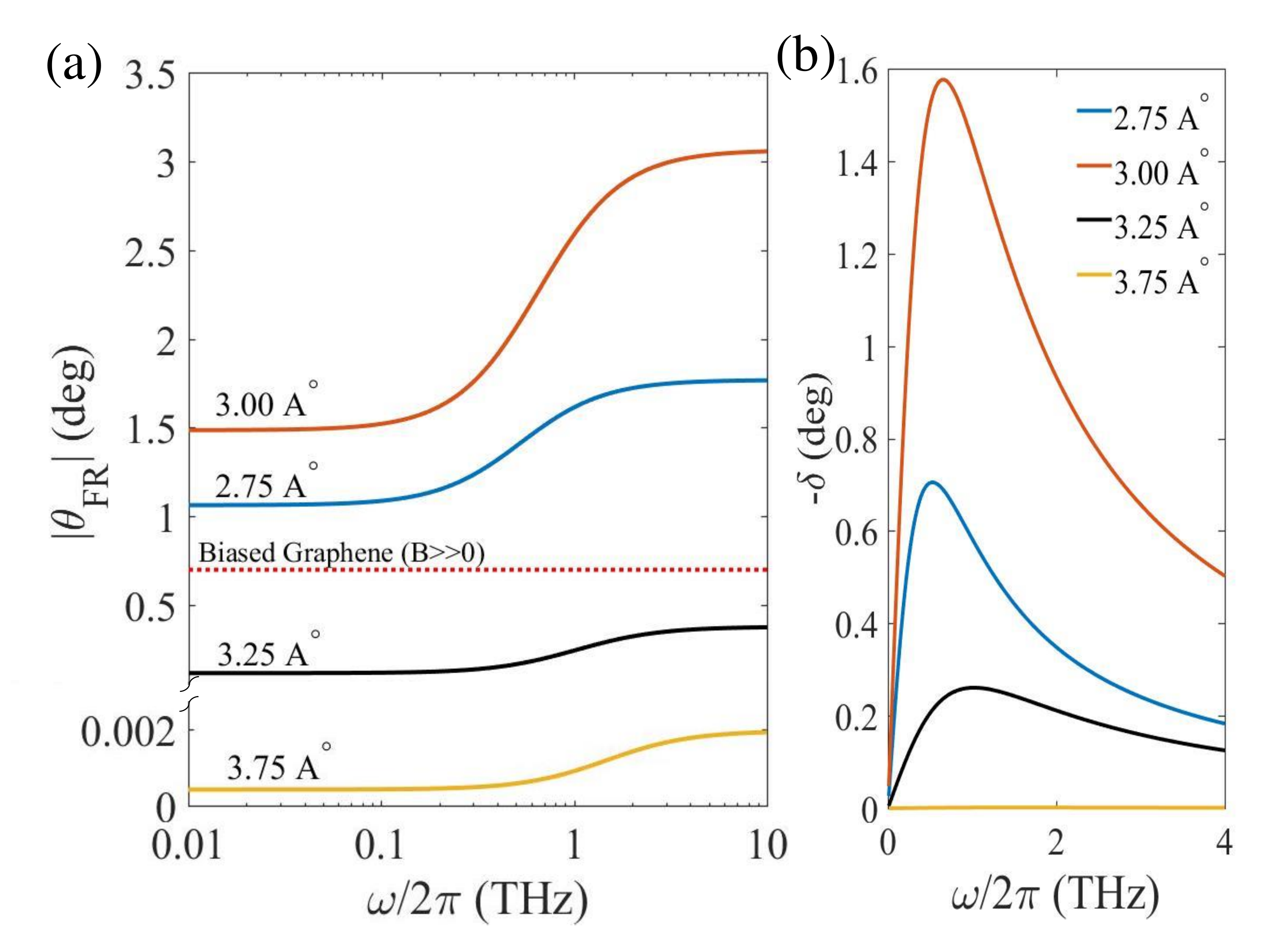}
		\caption{Faraday rotation (a) and ellipticity (b) of the CrI$_3$-graphene heterostructure as a function of frequency for different separations. The dotted  curve shows FR for externally-biased graphene using the effective field strengths from Fig. \ref{fig-mu}.} 
		\label{FRE}
\end{figure}

\begin{figure}[b]
		\includegraphics[width=1.0\columnwidth]{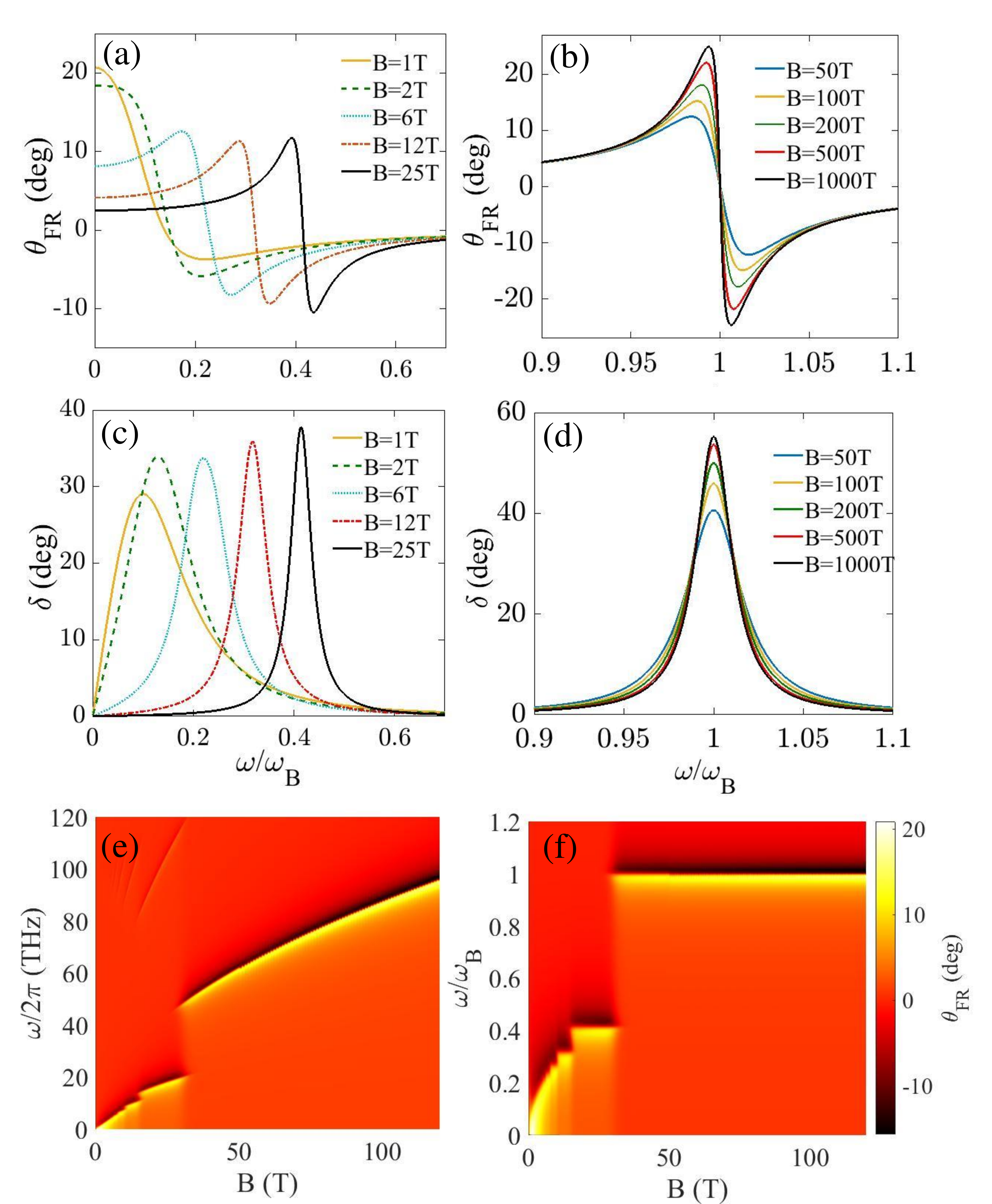}
		\caption{Faraday rotation (a)-(b) and ellipticity (c)-(d) of
graphene as a function of frequency for different external bias values,
and (e)-(f) density plot of Faraday rotation. $\omega_B$ is the
frequency of the first Landau level for each bias. $\mu=0.2$\,eV,
$T=40$\,K, and $\Gamma$=2 THz.} 
		\label{FR2}
\end{figure}

Faraday rotations for the exchange field case are shown in Fig.~\ref{FRE} for
various CrI$_3$-graphene separations. For closer-than-equilibrium separations,
modest Faraday rotations are observed, but for the equilibrium separation, FR is negligible since $\sigma_{xy}/\sigma_{xx}$ is small.
In contrast, for externally-biased graphene, Fig.~\ref{FR2},
large Faraday rotations can be obtained. As $B$ increases, the FR
resonance first blue-shifts (Fig.~\ref{FR2}(a)), and eventually stabilizes
in frequency (Fig.~\ref{FR2}(b)) at the first LL, but the peak FR
continues to increase with increasing $B$. Ellipticity behaves in a
similar manner as FR.  

\section{Conclusions}

We have examined exchange splitting in a monolayer chromium triiodide
(CrI$_3$)--graphene van der Waals (vdW) heterostructure using density-functional
theory where effective exchange fields of hundreds of Tesla are predicted. These
enormous fields serve as the magnetic bias for the graphene layer.
Graphene conductivity and SPP properties for the exchange field were considered, and
compared with the external bias case. Since no Landau levels occur for the exchange
field,  the resulting non-reciprocity is found to be considerably weaker than for an
equivalent external field bias (where strongly nonreciprocal electromagnetic edge modes
that are tightly-confined, robust, and unidirectional are shown to exist). Faraday rotation due to the exchange field was also shown to be modest compared to the external bias case.


\begin{thebibliography}{99}
\bibitem{CLB} In the $B=0$ case, the SPP is not formed for $\hslash \omega \gg \mu$ (states are not filled). In the external magnetic bias case, the value of $B$ plays a role, with larger $B$ moving the first LL to higher frequencies, increasing the bandgap.
\end{thebibliography}
\end{document}


\title{Supplemental Information for Exchange splitting and exchange-induced non-reciprocal photonic behavior of graphene in CrI$_3$-graphene vdW heterostructures}

\author{Alexander M. Holmes}
\email[]{holmesam@uwm.edu}
\affiliation{Department of Electrical Engineering, University of Wisconsin-Milwaukee, 3200 N. Cramer St., Milwaukee, Wisconsin 53211, USA}

\author{Samaneh Pakniyat}
\email[]{pakniyat@uwm.edu}
\affiliation{Department of Electrical Engineering, University of Wisconsin-Milwaukee, 3200 N. Cramer St., Milwaukee, Wisconsin 53211, USA}

\author{S. Ali Hassani Gangaraj}
\email[]{ali.gangaraj@gmail.com}
\affiliation{School of Electrical and Computer Engineering, Cornell University, Ithaca, NY 14853, USA}

\author{Francesco Monticone}
\email[]{francesco.monticone@cornell.edu}
\affiliation{School of Electrical and Computer Engineering, Cornell University, Ithaca, NY 14853, USA}

\author{George W. Hanson}
\email[]{george@uwm.edu}
\affiliation{Department of Electrical Engineering, University of Wisconsin-Milwaukee, 3200 N. Cramer St., Milwaukee, Wisconsin 53211, USA}

\author{Michael Weinert}
\email[]{weinert@uwm.edu}
\affiliation{Physics Department, University of Wisconsin-Milwaukee, Milwaukee, Wisconsin, 53211, USA}

\date{\today}

\maketitle

\section{Graphene Bands}

\subsection{Separation dependence}

\begin{figure}
 \includegraphics[width=\columnwidth]{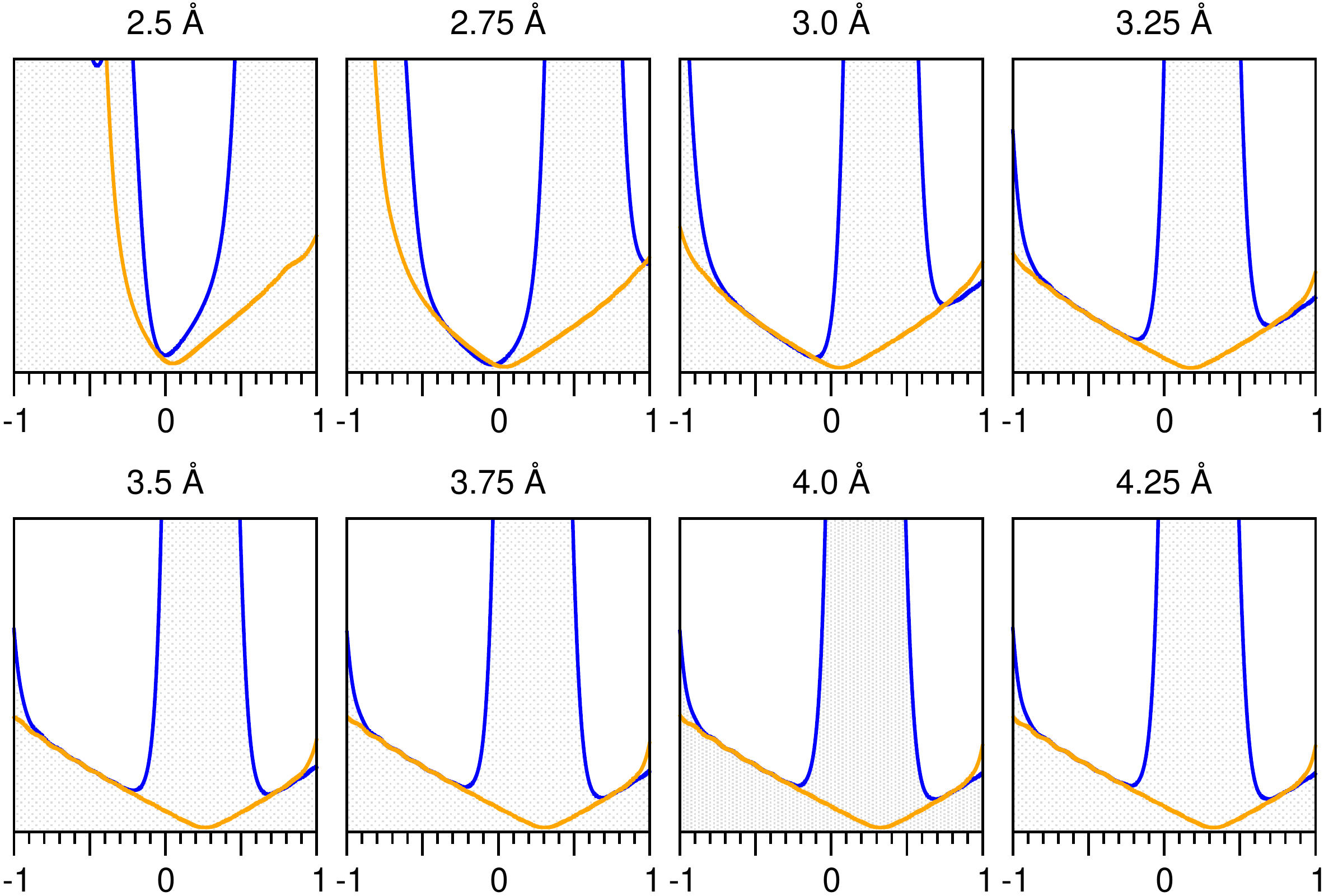}
 \caption{
 \label{fig:dos}
 Spin-resolved (majority: blue; minority: orange) density of states for the combined
graphene-CrI$_3$ for different separations obtained by broadening each eigenvalue by the
derivative of the Fermi function with $k_\mathrm{B}T$=0.040 eV. The Fermi level is set to
zero.
}
\end{figure}

The spin-resolved DOS around the Fermi level of the combined system as a function of
graphene-CrI$_3$ separation is shown in Fig.~S\ref{fig:dos}. The majority DOS has a large
peak above the Fermi level arising from the CrI$_3$ conduction bands, as well as the
graphene bands. For separations greater than $\sim$3.25 \AA, the minority DOS show the
characteristic V-shape due to the linear dispersion of the graphene bands around K, and
similarly the majority DOS show the linear behavior in the gaps on either side of the
CrI$_3$ peak. For closer approaches, however, the graphene contributions to the DOS of both
spins are significantly distorted despite the fact that the Dirac point lies in the CrI$_3$
band gap.

\begin{figure}[tb]
  \includegraphics[width=\columnwidth]{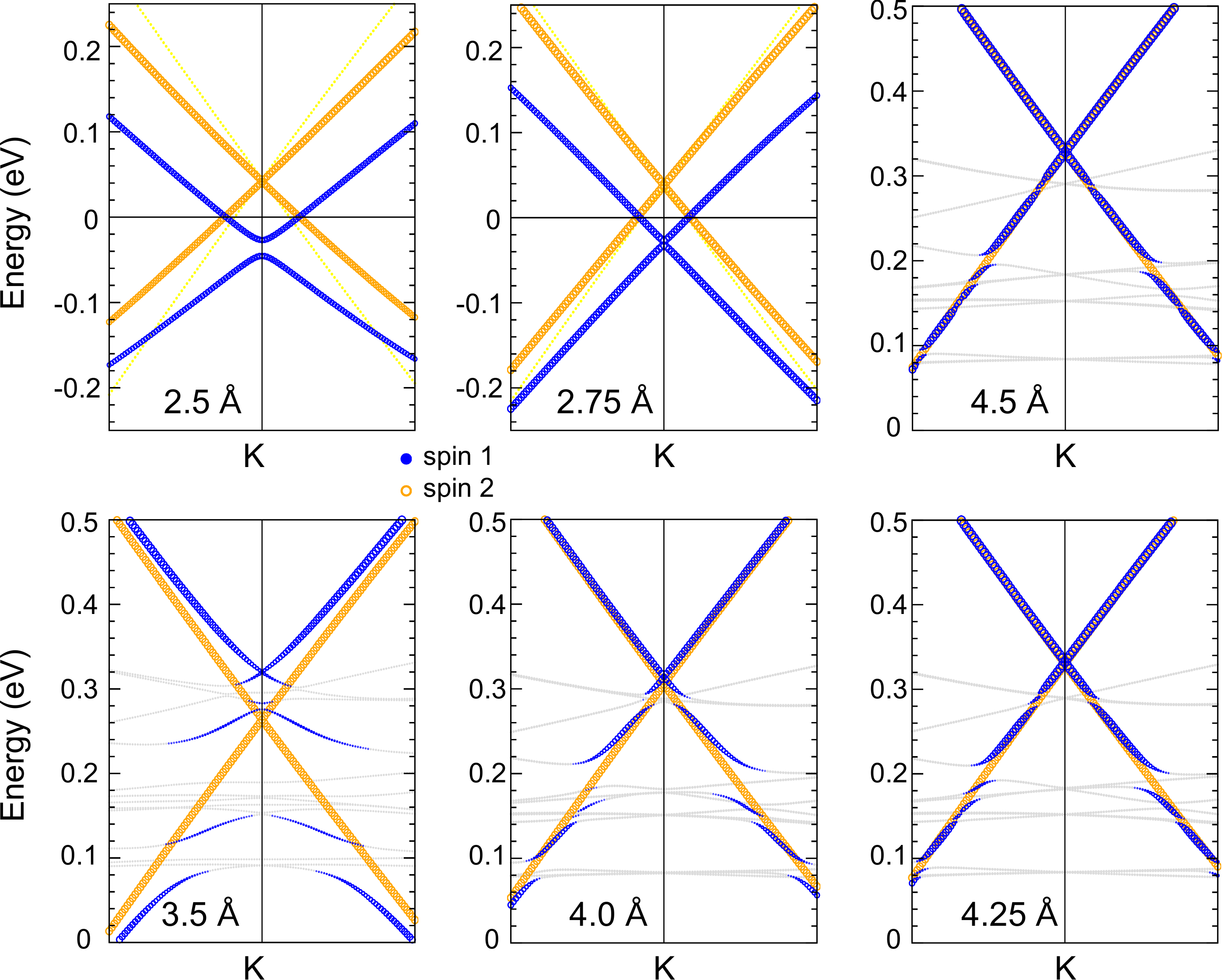}
  \caption{ $k$-projected bands of graphene around the K along ($\pm\frac{1}{40}$)
$\Gamma$-K of the (1$\times$1) Brillouin zone (c.f., Fig.~1 of the main
text) for different
graphene--CrI$_3$ separations.  Blue (orange) denote the majority
(minority) spin of the CrI$_3$, and the size of the circles represents the relative weight;
gray symbols denote bands with small graphene weight. The yellow lines in the 2.5 and 2.75
\AA{} plots are the isolated graphene bands shifted so the Dirac point coincides with that
of the minority bands.
  }
 \label{fig:bs-vs-z}
\end{figure}

The $k$-projected graphene bands for separations $\Delta d$ in addition to Fig.~2(c) of the
main text are shown in Fig.~S\ref{fig:bs-vs-z}. 
For  $\Delta d$=2.5 \AA,
the Fermi level is in the CrI$_3$ gap.
The bands are strongly exchanged split and, in order to maintain charge
neutrality, the majority states are electron-doped while the minority states are
hole-doped. The majority -- but not the minority -- bands are strongly
split, 18.5 meV, at
the Dirac point due to interactions with the CrI$_3$ substrate. 
(The minority Dirac point has a small 0.3 meV gap.) This different behavior,
which is noticeable at all separations,
is a consequence of the fact that the nearby CrI$_3$ states
are of majority character. In addition, although the dispersion of the minority bands is
linear, the slope is noticeably smaller, with the difference decreasing as the separation
increases.

As the graphene and CrI$_3$ move apart, the Fermi level moves into the CrI$_3$
(majority spin) conduction band. As a consequence, the dispersions of the majority states below the Dirac
point are modified via interactions and hybridization with the substrate
bands, forming mini-bands in the graphene and CrI$_3$
bands. These mini-bands continue to exist even to quite large separations (e.g., 4.25 \AA),
and thus will affect the optical properties..

\subsection{Spin-orbit effects}

\begin{figure}[t]
  \includegraphics[width=0.85\columnwidth]{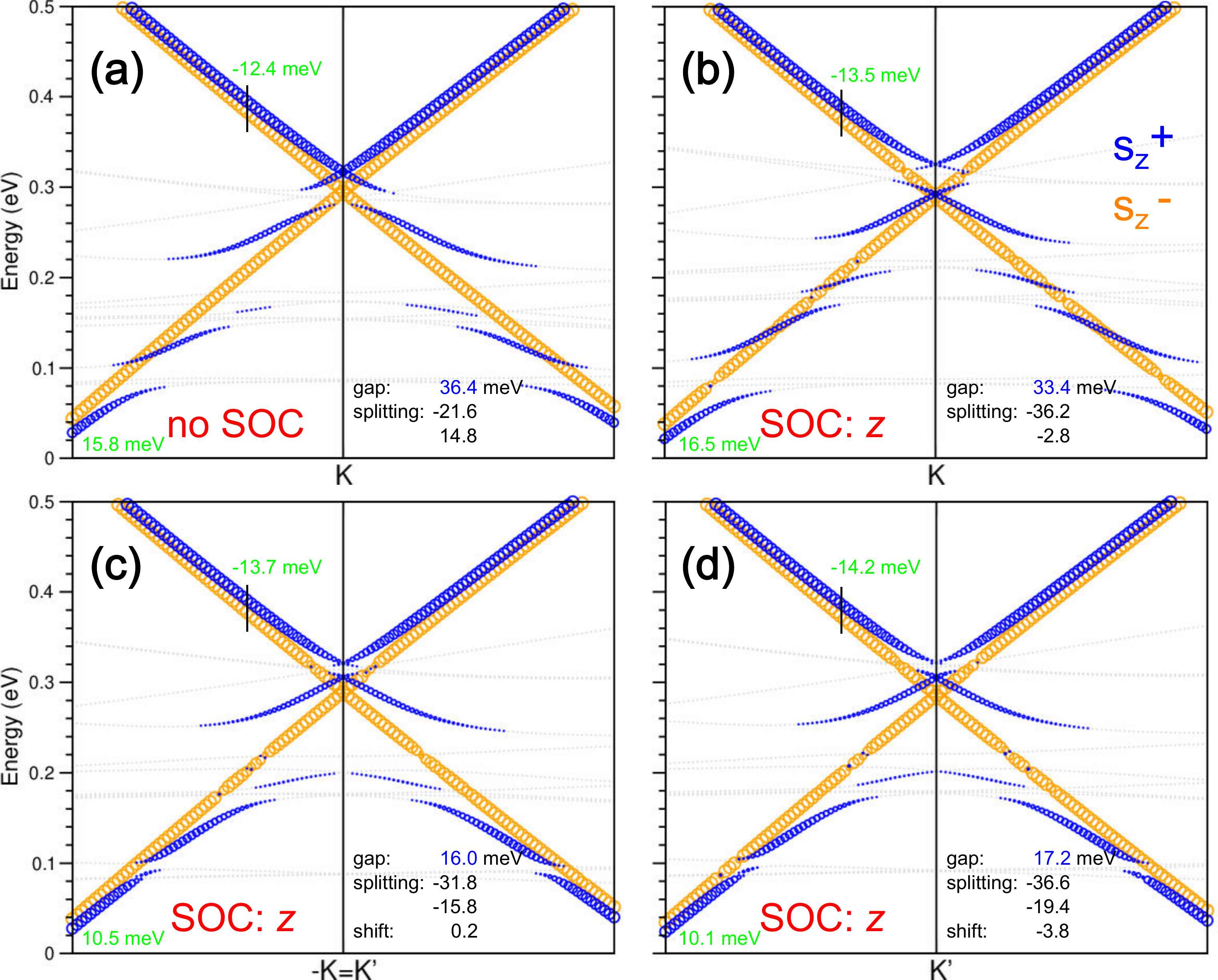}
  \caption{
 $k$-projected bands of graphene around the various K points of the (1$\times$1) Brillouin zone as in
Fig.~1(b) of the main text ($\pm\frac{1}{40}$ along
$\Gamma$-K) for the
equilibrium separation of 3.75 \AA{}: (a) No spin-orbit; and including
spin-orbit with the Cr magnetic moments
oriented along the \textit{z}-axis: for (b) K, (c) $-$K=K', and (d) the K'
rotated by 60$^o$. Blue (orange) symbols denote majority (minority)
spins in (a) and projection along $\pm\hat{z}$ in (b)-(d), with the size
corresponding to relative weight; gray symbols are for are states with
small projected weights. The exchange splittings away from the Dirac
points are given in green; for the Dirac points,``gap'' refers to the
splitting in the majority/$s_z^+$ bands, the two values of
``splittings'' are with respect to the Dirac point of the
minority/$s_z^-$ states, and ``shift'' in (c) and (d) is relative (b).
  }
 \label{fig:soc-z}
\end{figure}

Without spin-orbit, the majority and minority bands separately are
symmetric under the transformation $\vec{k}\to -\vec{k}$, i.e., the
states at K and $-$K=K are degenerate. For a system with spin-orbit and
magnetism (broken time-reversal symmetry), however, the these
degeneracies no longer need to hold. In Fig.~S\ref{fig:soc-z} the
projected bands are shown for moments oriented along the
\textit{z}-axis, the calculated preferred direction. The bands at K,
Fig.~S\ref{fig:soc-z}(b) differ from those at $-$K  and K',
Fig.~S\ref{fig:soc-z}(c,d), in the splittings and in the size and shape
of the mini-bands. (There are also smaller differences between $-$K and
K'.) These differences open the possibility of non-reciprocal
valleytronic effects.

\begin{figure}[t]
  \includegraphics[width=0.85\columnwidth]{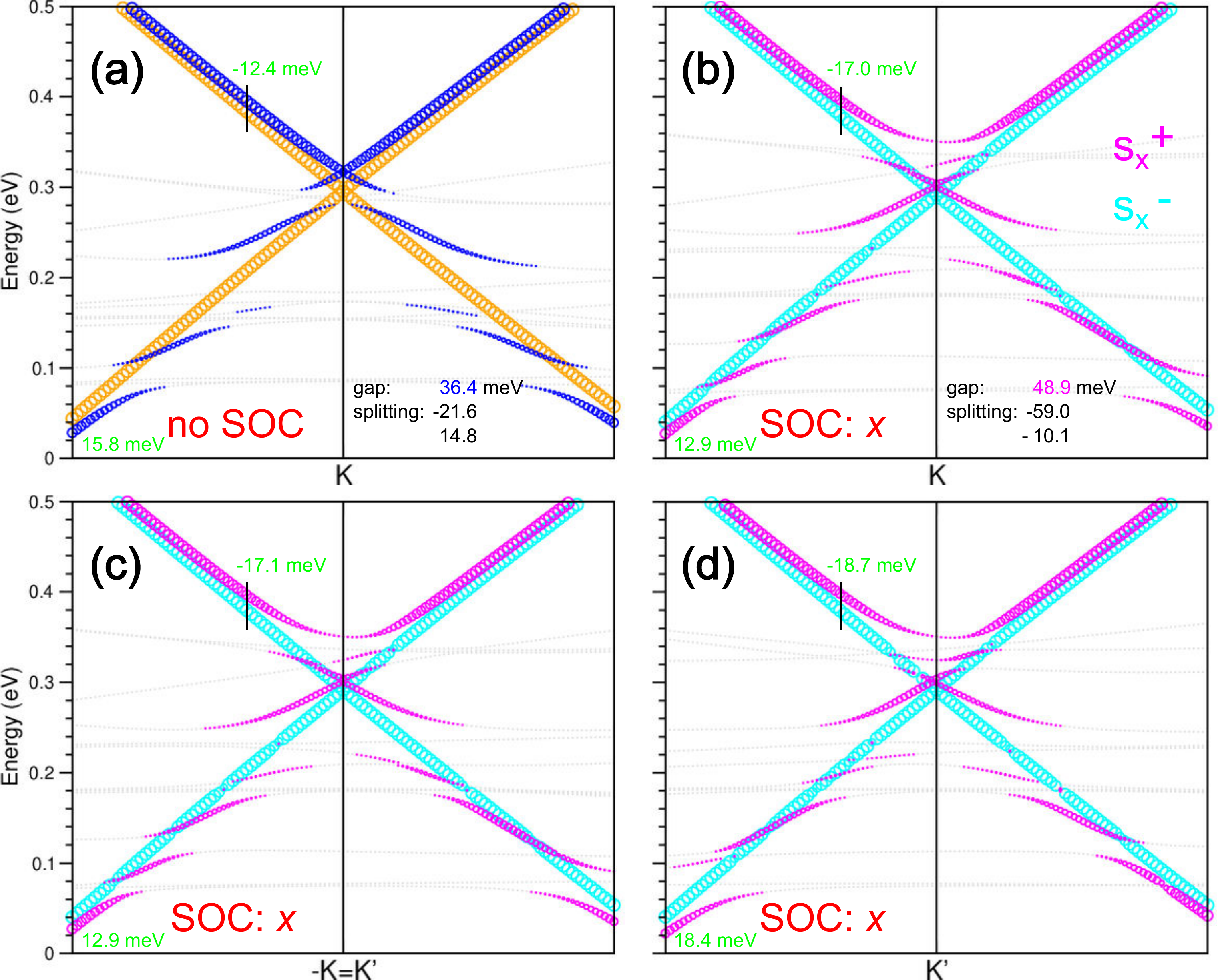}
  \caption{As in Fig.~S\ref{fig:soc-z} except the Cr moments are aligned
along the \textit{x}-axis.
  }
 \label{fig:soc-x}
\end{figure}

The direction of the magnetic moments causes noticeable effects on the
dispersions. The results for moments oriented along the \textit{x}-axis
are shown in Fig.~S\ref{fig:soc-x}. As before, the graphene bands are
essentially fully polarized along the axis of the Cr moments. The
various splittings and the mini-bands change compared  to those in for
the moments along the \textit{z}-axis.

\subsection{Optical Conductivity}

\begin{figure}[tb]
 \includegraphics[width=0.95\columnwidth]{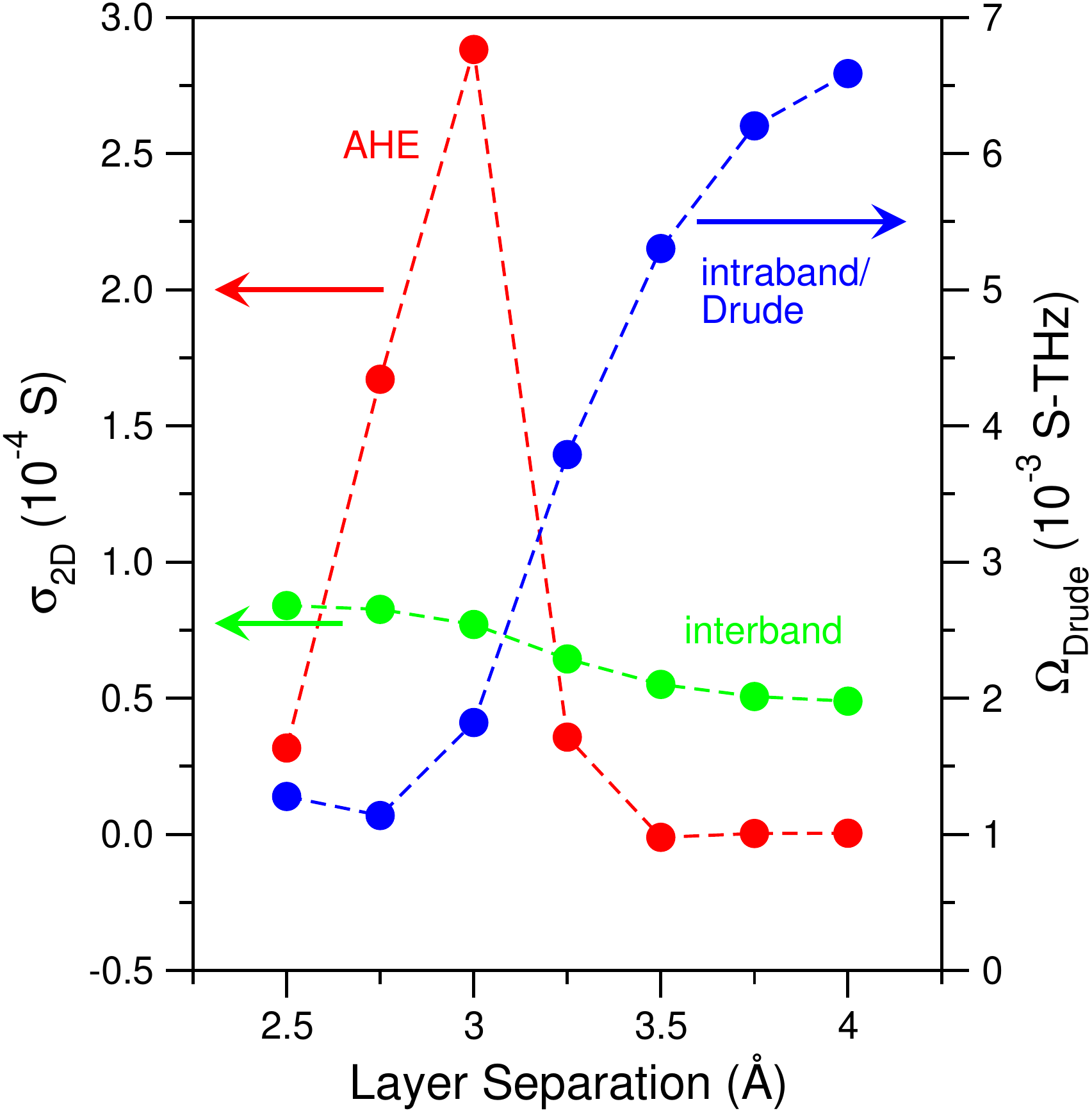}
 \caption{
 \label{fig:sigma}
 Calculated 2D contributions to the optical properties: Drude intraband contribution,
$\Omega_\mathrm{Drude}$ (blue); anomalous Hall, $\sigma_{xy}^{2D}$ (red); and interband,
$\sigma_{xx}^{2D}$ (green).
 }
\end{figure}

The 2D optical conductivity
for graphene-CrI$_3$ and isolated graphene were
calculated
within the independent particle approximation \cite{vasp-opts} and shown in
Fig.~S\ref{fig:sigma}. 
For graphene, with its linearly dispersing bands,
the real part of the interband contribution of $\sigma_\mathrm{2D}^{xx}(\omega)$ goes to
$\frac{e^2}{4\hbar}$ for low frequencies; the calculated conductivity
correctly obeys this limit, suggesting that the computational
parameters (particularly $k$-point sampling) are adequate for the present purposes. For the
composite system, the interband contribution 

The intraband Drude-like contribution to the 2D conductivity tensor,
$\sigma_\mathrm{2D}^\mathrm{Drude}$, is given in terms of the plasma
frequency tensor $\omega_p^2$ 
\begin{equation}
 \left(\omega_p^2\right)_{\alpha\beta} \sim \sum_{n\bm{k}} f_{n\bm{k}}
\frac{\partial^2\epsilon_{n\bm{k}}}{\partial\bm{k}_\alpha \partial\bm{k}_\beta}
\end{equation}
by
\begin{eqnarray}
 \sigma_\mathrm{2D}^\mathrm{Drude} & = &\frac{L}{4\pi} \,
 \frac{(\hbar\omega_p)^2}{(\hbar\omega)^2+(\hbar\gamma)^2} \,
    ( \gamma + i \omega )  \\
 & \equiv & \Omega_\mathrm{Drude} \,
  \frac{\gamma_\nu + i \nu}{\nu^2+\gamma_\nu^2} ,
\end{eqnarray}
where $\omega=2\pi\nu$ and $\gamma_\nu=2\pi\gamma$. As expected, $\Omega_\mathrm{Drude}$
increases as the as the Fermi level crosses into the CrI$_3$ conduction band. (Only the
$xx$ component is appreciable.) For reasonable values of $\hbar\gamma$ of a few THz, the
intraband dominates the interband contribution to $\sigma_{xx}$.

Because of the exchange field due to the CrI$_3$, there is a small contribution to
$\sigma_{xy}$, but the largest contribution (within the present approximations) comes from
the anomalous Hall effect (AHE),
\begin{equation}
 \sigma^\mathrm{AHE}_{xy} \sim \mathrm{Im} \sum_{nm} (f_n-f_m)
 \braket{u_m}{\frac{d}{dk_x} u_n} \braket{u_m}{\frac{d}{dk_y} u_n}^* \ .
\end{equation}
This term grows, and is large, as the graphene Dirac point and Fermi level approach the bottom of the
CrI$_3$ conduction band, and then a sudden collapse as the Fermi level crosses into the
CrI$_3$. Thus, for the equilibrium spacing, this contribution is again small. If the large
value of $\sigma_{xy}$ is due to the proximity of the Fermi level near the bottom of the
CrI$_3$ conduction band -- and the existence of mini-bands in the graphene majority bands -- it is possible that a GW or hybrid functional calculation that
increases the gap might show a large AHE contribution at the equilibrium separation.

\subsection {Graphene Conductivity in an External Bias}
For the graphene conductivity in an external magnetic bias, we use the expressions in \cite{Gus1}. A prominent feature in strong external bias fields is the occurrence of Landau levels, whereas, as described in the main text, LLs are absent for the exchange field. It should be noted that in \cite{Gus1} spin splitting is ignored, so that summation over spin states just gives a multiplicative factor of 2. Here, spin splitting can be significant. At the equilibrium separation between graphene and CrI\textsubscript{3}, the effective chemical potentials of the two spins are similar in magnitude, and have the same sign (main text, Fig.~3b). In this case, summing the conductivity in \cite{Gus1} over these two spins gives approximately the same result as the simple factor of 2. In the event of a strained system having smaller separation, the chemical potentials of the two spins can have opposite sign (main text, Fig.~3b). In that case, the two spin contributions can partially cancel, decreasing the magneto-optical conductivity terms.

\section{Edge Mode Model I: Electrostatic Potential}

The following derivation loosely follows \cite{Apell}. We also obtain a new approximate dispersion solution for the fundamental edge mode. 

\subsection{Electrostatic Green Function for a Homogeneous Single Interface Structure}

\begin{figure}[b]
		\includegraphics[clip,width=0.65\columnwidth]{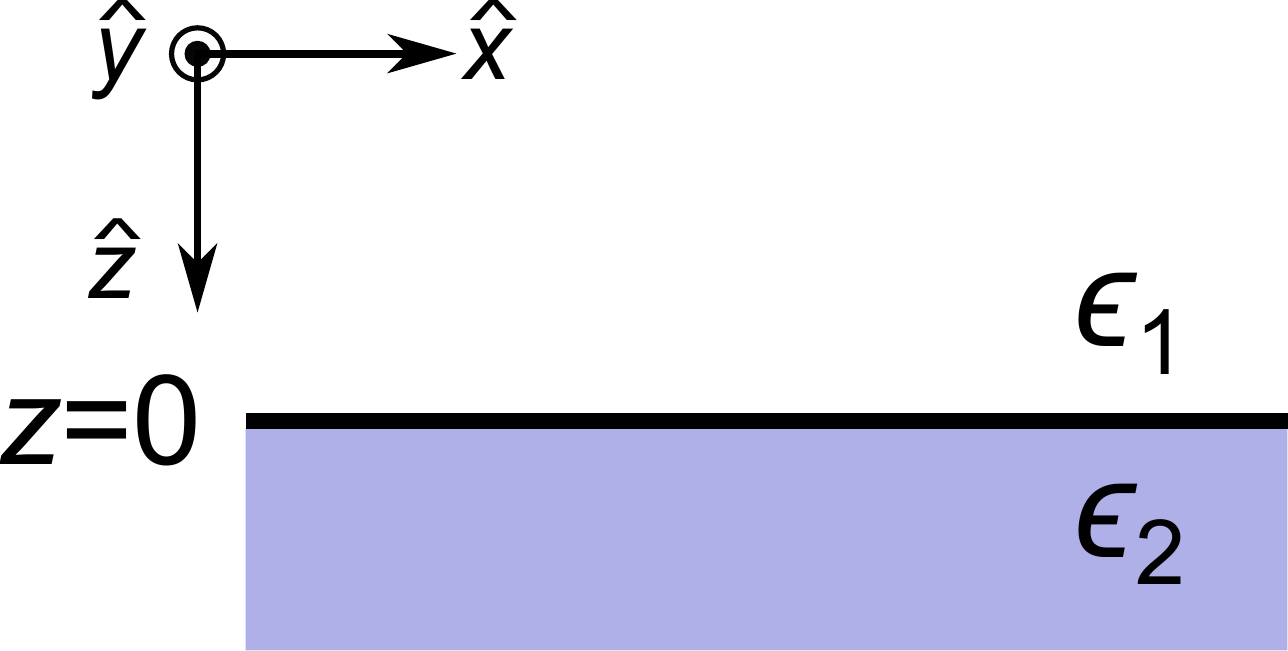}
		\caption{Schematic of a two layer, laterally-infinite, dielectric medium.} 
		\label{interface}
\end{figure}

Consider the dielectric interface structure in Fig.~S\ref{interface}.
For a charge distribution assumed to be in Region I ($z<0$), Poisson's
equation relates the electrostatic potential to the net charge density as%
\begin{equation}
\nabla ^{2}\left\{ 
\begin{array}{c}
\Phi _{1}\left( \mathbf{r}\right) \\ 
\Phi _{2}\left( \mathbf{r}\right)%
\end{array}%
\right\} =\left\{ 
\begin{array}{c}
-\rho \left( \mathbf{r}\right) /\varepsilon _{1} \\ 
0%
\end{array}%
\right\} ,  \label{re1}
\end{equation}
subject to the boundary conditions at the interface%
\begin{align}
\left. \Phi _{1}\left( \mathbf{r}\right) \right\vert _{z=0}& =\left. \Phi
_{2}\left( \mathbf{r}\right) \right\vert _{z=0},  \label{re2} \\
\varepsilon _{1}\left. \frac{\partial }{\partial z}\Phi _{1}\left( \mathbf{r}%
\right) \right\vert _{z=0}& =\varepsilon _{2}\left. \frac{\partial }{%
\partial z}\Phi _{2}\left( \mathbf{r}\right) \right\vert _{z=0}\ .
\label{re3}
\end{align}
The associated Green function for each region then satisfies%
\begin{equation}
\nabla ^{2}\left\{ 
\begin{array}{c}
G_{1}\left( \mathbf{r},\mathbf{r}^{\prime }\right) \\ 
G_{2}\left( \mathbf{r},\mathbf{r}^{\prime }\right)%
\end{array}%
\right\} =\left\{ 
\begin{array}{c}
-\delta \left( \mathbf{r}-\mathbf{r}^{\prime }\right) /\varepsilon _{1} \\ 
0%
\end{array}%
\right\} ,  \label{re4}
\end{equation}
subject to the same boundary conditions. The electrostatic potential in the
$i$th region may then be written as%
\begin{equation}
\Phi _{i}\left( \mathbf{r}\right) =\int G_{i}\left( \mathbf{r},\mathbf{r}%
^{\prime }\right) \rho \left( \mathbf{r}^{\prime }\right) d^{3}\mathbf{r}%
^{\prime }\ .  \label{re5}
\end{equation}

In the spacial transform domain, it is easy to show that the particular
solution of (\ref{re4}) in Region I is%
\begin{equation}
G_{1}^{p}\left( \mathbf{k},\mathbf{r}^{\prime }\right) =G_{1}^{p}\left( 
\mathbf{k}\right) e^{-i\mathbf{k}\cdot \mathbf{r}^{\prime }} ,  \label{re6}
\end{equation}
where $G_{1}^{P}\left( \mathbf{k}\right) =1/\varepsilon _{1}\left\vert 
\mathbf{k}\right\vert ^{2}$. The principal Green function in
Region I, is then given by the inverse spacial transform of (\ref{re6}) with
respect to $z$, 
\begin{align}
G_{1}^{p}\left( \mathbf{q},z,z^{\prime }\right) & =\frac{1}{2\pi }%
\int_{-\infty }^{\infty }dk_{z}G^{p}\left( \mathbf{k}\right) e^{ik_{z}\left(
z-z^{\prime }\right) } \\
& =\frac{1}{2\varepsilon _{1}q}e^{-q\left\vert z-z^{\prime }\right\vert } ,
\label{re8}
\end{align}
where $\mathbf{q}\equiv \mathbf{\hat{x}}k_{x}+\mathbf{\hat{y}}k_{y}$ and $%
q\equiv \left\vert \left\vert \mathbf{q}\right\vert \right\vert =\sqrt{%
k_{x}^{2}+k_{y}^{2}}$. In addition to the principal Green function in
Region I, we add a homogeneous contribution $G_{1}^{H}\left( \mathbf{q}%
,z\right) $ satisfying $\nabla ^{2}G_{1}^{H}=0$. Since there is no source
terms in Region II, the Green function there consists only of a homogeneous term, $%
G_{1}\left( \mathbf{q},z,z^{\prime }\right) =G_{2}^{H}\left( \mathbf{q}%
,z\right) $ where $G_{2}^{H}\left( \mathbf{q},z\right) $ satisfies $\nabla
^{2}G_{2}^{H}=0$. It is easy to show that%
\begin{eqnarray}
G_{1}^{H} &=&A\left( \mathbf{q}\right) e^{qz} , \label{re9} \\
G_{2}^{H} &=&B\left( \mathbf{q}\right) e^{-qz} , \label{re10}
\end{eqnarray}
where $A\left( \mathbf{q}\right) $ and $B\left( \mathbf{q}\right) $ are
determined by applying the boundary conditions (\ref{re2})-(\ref{re3}) at
the interface. For $z^{\prime }\leq 0$, it follows that%
\begin{align}
G_{1}\left( \mathbf{q},z,z^{\prime }\right) & =G_{1}^{p}\left( \mathbf{q}%
,z,z^{\prime }\right) +A\left( \mathbf{q}\right) e^{qz},  \label{re58} \\
G_{2}\left( \mathbf{q},z,z^{\prime }\right) & =B\left( \mathbf{q}\right)
e^{-qz} ,  \label{re59}
\end{align}
where%
\begin{align}
A\left( \mathbf{q}\right) & =\frac{1}{2\varepsilon _{1}}\frac{\varepsilon
_{1}-\varepsilon _{2}}{\varepsilon _{1}+\varepsilon _{2}}\frac{e^{qz^{\prime
}}}{q},  \label{re11} \\
B\left( \mathbf{q}\right) & =\frac{1}{\varepsilon _{1}+\varepsilon _{2}}%
\frac{e^{qz^{\prime }}}{q} .  \label{re12}
\end{align}
In the limit $z^{\prime }\rightarrow 0$, we obtain the Green function for
a source positioned at the interface,%
\begin{equation}
G\left( \mathbf{q},z,0\right) =\frac{1}{2\bar{\varepsilon}}\frac{%
e^{-q\left\vert z\right\vert }}{q},  \label{re13}
\end{equation}
where $\bar{\varepsilon}\equiv \left( \varepsilon _{1}+\varepsilon
_{2}\right) /2$. This Green function accounts for the background structure
that will host the graphene.

\subsection{Charge Density on Semi-Infinite Graphene}

In this section, we consider a 2D charge density on graphene localized at $z=0$. The graphene exists for $x <0$, invariant with respect to $y$. The charge density is given by%
\begin{equation}
\rho \left( \mathbf{r}\right) =\rho _{s}\left( x\right) \delta \left(
z\right) e^{ik_{y}y},  \label{re14}
\end{equation}
where $\rho _{s}\left( x\right) $ denotes the surface charge density at the
interface. Because the electrostatic potential is also invariant with
respect to $y$, we write $\Phi \left( \mathbf{r}\right) =\Phi \left(
x,z\right) e^{ik_{y}y}$. Application of (\ref{re5}) leads to%
\begin{equation}
\Phi \left( x,z\right) =\int_{-\infty }^{\infty }dx^{\prime }G\left(
x,x^{\prime },z,0\right) \rho _{s}\left( x^{\prime }\right) ,  \label{re15}
\end{equation}
where%
\begin{align}
G\left( x,x^{\prime },z,0\right) & =\frac{1}{2\bar{\varepsilon}}%
\int_{-\infty }^{\infty }\frac{dk_{x}}{2\pi }\frac{e^{-q\left\vert
z\right\vert }}{q}e^{ik_{x}\left( x-x^{\prime }\right) } \\
& =\frac{1}{2\pi \bar{\varepsilon}}K_{0}\left( \left\vert k_{y}\right\vert 
\sqrt{\left( x-x^{\prime }\right) ^{2}+z^{2}}\right) ,  \label{re16}
\end{align}
with $K_{0}$ denoting the zero-order modified Bessel function of the second
kind. The absolute value $\left\vert k_{y}\right\vert$ arises from having $q =\sqrt{k_{x}^{2}+k_{y}^{2}}$.

The continuity equation relates the surface charge density to the surface
current at the interface by $i\omega \rho _{s}\left( x\right) =\mathbf{%
\nabla }\cdot \mathbf{J}_{s}\left( x\right) $ where $\mathbf{J}_{s}\left(
x\right) =\Theta \left( -x\right) \mathbf{\bar{\sigma}}\cdot \left. -\mathbf{%
\nabla }\Phi \left( x,z\right) \right\vert _{z=0}$. The components of the
current expand to%
\begin{align}
J_{sx}\left( x\right) & =-\Theta \left( -x\right) \left[ ik_{y}\sigma
_{xy}+\sigma _{xx}\frac{d}{dx}\right] \Phi \left( x,0\right) ,
\label{re18} \\
J_{sy}\left( x\right) & =-\Theta \left( -x\right) \left[ ik_{y}\sigma
_{yy}+\sigma _{yx}\frac{d}{dx}\right] \Phi \left( x,0\right) ,
\label{re19}
\end{align}
which are used in the continuity equation to obtain $\rho _{s}\left(
x\right) \equiv \delta \left( -x\right) \rho _{e}\left( x\right) +\Theta
\left( -x\right) \rho _{b}\left( x\right) $, where%
\begin{align}
\rho _{e}\left( x\right) & \equiv \hat{D}_{e}\left( x\right) \Phi \left(
x,0\right) ,  \label{re20} \\
\rho _{b}\left( x\right) & \equiv \hat{D}_{b}\left( x\right) \Phi \left(
x,0\right) ,  \label{re21}
\end{align}
such that%
\begin{align}
\hat{D}_{e}\left( x\right) & \equiv k_{y}\chi _{xy}+\eta _{xx}\frac{d}{dx},
\label{re22} \\
\hat{D}_{b}\left( x\right) & \equiv k_{y}^{2}\eta _{yy}-k_{y}\left( \chi
_{xy}+\chi _{yx}\right) \frac{d}{dx}-\eta _{xx}\frac{d^{2}}{dx^{2}},
\label{re23}
\end{align}
where we define $\eta _{\alpha \alpha }\equiv \sigma _{\alpha \alpha
}/i\omega $ and $\chi _{\alpha \beta }\equiv \sigma _{\alpha \beta }/\omega $
for $\alpha ,\beta \in \left\{ x,y\right\} $. Substituting (\ref{re20})-(\ref%
{re21}) into (\ref{re5}), we have%
\begin{align}
\Phi \left( x,z\right) & =\int_{-\infty }^{\infty }G\left( x,x^{\prime
},z,0\right) \rho _{s}\left( x^{\prime }\right) \\
& =G\left( x,0,z,0\right) \rho _{e}\left( 0\right) +\int_{-\infty
}^{0}dx^{\prime }G\left( x,x^{\prime },z,0\right) \rho _{b}\left( x^{\prime
}\right) ,  \label{re24}
\end{align}
where $\rho _{e}\left( 0\right) $ and $\rho _{b}\left( x\right) $ should be
interpreted as the charge density at the edge ($x=0$) and in the bulk region
($x<0$), respectively. Setting $z=0$, we obtain an integro-differential
equation for the potential in the plane of the interface,%
\begin{equation}
\phi \left( x\right) =g\left( x,0\right) \rho _{e}\left( 0\right)
+\int_{-\infty }^{0}dx^{\prime }g\left( x,x^{\prime }\right) \rho _{b}\left(
x^{\prime }\right) ,  \label{re25}
\end{equation}
where $\phi \left( x\right) \equiv \Phi \left( x,0\right) $ and $g\left(
x,x^{\prime }\right) \equiv G\left( x,x^{\prime },0,0\right) $.

We now expand the potential in terms of Laguerre polynomials,%
\begin{equation}
\phi \left( x\right) =e^{\left\vert k_{y}\right\vert x}\sum_{n=0}^{\infty
}c_{n}L_{n}\left( -2\left\vert k_{y}\right\vert x\right) ,  \label{re26}
\end{equation}
for which we have the orthogonality condition%
\begin{equation}
\int_{-\infty }^{0}e^{2\left\vert k_{y}\right\vert x}L_{m}\left(
-2\left\vert k_{y}\right\vert x\right) L_{n}\left( -2\left\vert
k_{y}\right\vert x\right) dx=\frac{\delta _{nm}}{2\left\vert
k_{y}\right\vert }\ .  \label{re27}
\end{equation}

Exploiting orthogonality by multiplying both sides sides of (\ref{re25}) by $%
e^{\left\vert k_{y}\right\vert x}L_{m}\left( -2\left\vert k_{y}\right\vert
x\right) $ and integrating over $x$ from $0$ to $\infty $ leads to the
dispersion relation%
\begin{equation}
\frac{c_{m}}{2\left\vert k_{y}\right\vert }=\sum_{n=0}^{\infty }c_{n}A_{mn}\
,  \label{re28}
\end{equation}
where%
\begin{equation}
A_{mn}\equiv J_{mn}+\eta _{xx}\left( 2n+1\right) I_{m}+\text{sgn}\left(
q_{y}\right) \chi _{xy}I_{m},  \label{re60}
\end{equation}
with%
\begin{align}
I_{m}& \equiv \left\vert k_{y}\right\vert \int_{-\infty }^{0}dxe^{\left\vert
k_{y}\right\vert x}L_{m}\left( -2\left\vert k_{y}\right\vert x\right)
g\left( x,0\right) ,  \label{re29} \\
J_{mn}& \equiv \int_{-\infty }^{0}\int_{-\infty }^{0}dxdx^{\prime }G\left(
x,x^{\prime }\right) \hat{D}_{b}\left( x^{\prime }\right) e^{\left\vert
k_{y}\right\vert \left( x+x^{\prime }\right) } \label{re30} \\ 
&  \qquad \qquad \qquad  \times L_{m}\left( -2\left\vert
k_{y}\right\vert x\right) L_{n}\left( -2\left\vert k_{y}\right\vert
x^{\prime }\right)  . \notag  
\end{align}

Making the change of variable $y\equiv \left\vert k_{y}\right\vert x$,
reduces (\ref{re29})-(\ref{re30}) to%
\begin{align}
I_{m}& =\int_{-\infty }^{0}dyG\left( y,0\right) e^{y}L_{m}\left( -2y\right)
,  \label{re31} \\
J_{mn}& =\int_{-\infty }^{0}\int_{-\infty }^{0}dydy^{\prime }G\left(
y,y^{\prime }\right) \hat{D}_{b}\left( y^{\prime }\right) e^{\left(
y+y^{\prime }\right) } \label{re32}\\ 
&  \qquad \qquad \qquad  \times L_{m}\left( -2y\right) L_{n}\left( -2y^{\prime
}\right) .  \notag
\end{align}
where%
\begin{align}
G\left( y,y^{\prime }\right) & =\frac{1}{2\pi \bar{\varepsilon}}K_{0}\left(
\left\vert y-y^{\prime }\right\vert \right) ,  \label{re33} \\
\hat{D}_{b}\left( y^{\prime }\right) & =\eta _{yy}-\text{sgn}\left(
q_{y}\right) \left( \chi _{xy}+\chi _{yx}\right) \frac{d}{dy}-\eta _{xx}%
\frac{d^{2}}{dy^{2}}\ .  \label{re34}
\end{align}

Note that in this work, we assume $\sigma _{xx}=\sigma _{yy}$ and $\sigma
_{xy}=-\sigma _{yx}$, which significantly reduces (\ref{re32}) to%
\begin{equation}
J_{mn}=-\eta _{xx}\hat{S}\left( y,y^{\prime }\right) G\left( y,y^{\prime
}\right) e^{\left( y+y^{\prime }\right) }L_{m}\left( -2y\right) \frac{d^{2}}{%
dy^{\prime 2}}L_{n+1}\left( -2y^{\prime }\right) ,  \label{re35}
\end{equation}
which is straight forward to derive obtain using the recursive formulas
\begin{align}
e^{-y^{\prime }}\frac{d}{dy^{\prime }}e^{y^{\prime }}L_{n}\left( -2y^{\prime
}\right) & =\frac{d}{dy^{\prime }}L_{n+1}\left( -2y^{\prime }\right)
-L_{n}\left( -2y^{\prime }\right) ,  \label{re36} \\
e^{-y^{\prime }}\frac{d^{2}}{dy^{\prime 2}}e^{y^{\prime }}L_{n}\left(
-2y^{\prime }\right) & =\frac{d^{2}}{dy^{\prime 2}}L_{n+1}\left( -2y^{\prime
}\right) +L_{n}\left( -2y^{\prime }\right)  .  \label{re37}
\end{align}

Truncating the expansion to $N+1$ terms allows us to cast (\ref{re28}) as a
standard eigenvalue equation%
\begin{equation}
\left[ 
\begin{array}{cccc}
A_{00} & A_{01} & \cdots & A_{0N} \\ 
A_{10} & A_{11} & \cdots & A_{1N} \\ 
\vdots & \vdots & \ddots & \vdots \\ 
A_{N0} & A_{N1} & \cdots & A_{NN}%
\end{array}%
\right] \left[ 
\begin{array}{c}
c_{0} \\ 
c_{1} \\ 
\vdots \\ 
c_{N}%
\end{array}%
\right] =\lambda \left[ 
\begin{array}{c}
c_{0} \\ 
c_{1} \\ 
\vdots \\ 
c_{N}%
\end{array}%
\right] ,  \label{re38}
\end{equation}
where $\lambda \equiv 1/2\left\vert k_{y}\right\vert $.

\subsection{Surface Charge Density}

Once the eigenvalue equation is solved for $\left\{ c_{n}\right\} $, one can
obtain numerical solutions for the potential and the surface charge density.
Using (\ref{re20}),\ (\ref{re21}), and (\ref{re26}), it can be shown that
\begin{align}
\rho _{e}\left( y\right) & =\left\vert k_{y}\right\vert
e^{y}\sum_{n=0}^{\infty }c_{n}\left[ \left( \text{sgn}\left( k_{y}\right)
\chi _{xy}-\eta _{xx}\right) L_{n}\left( -2y\right) \right.   \nonumber \\
& \qquad \qquad \qquad \left. +\eta _{xx}\frac{d}{dy}L_{n+1}\left(
-2y\right) \right]   \label{re51} \\
\rho _{b}\left( y\right) & =k_{y}^{2}e^{y}\sum_{n=0}^{\infty }c_{n}\left[
\left( \eta _{yy}-\eta _{xx}+\text{sgn}\left( q_{y}\right) \left( \chi
_{xy}+\chi _{yx}\right) \right) \right.  \nonumber \\
& \qquad \times L_{n}\left( -2y\right) -\text{sgn}\left( q_{y}\right) \left(
\chi _{xy}+\chi _{yx}\right) \frac{d}{dy}L_{n+1}\left( -2y\right) \nonumber \\
& \qquad \left. -\eta _{xx}\frac{d^{2}}{dy^{2}}L_{n+1}\left( -2y\right) \right]  \label{re50}
\end{align}
Then, using $L_{n}\left( 0\right) =1$ and $L_{n}^{\prime }\left( 0\right) =-n
$, we obtain%
\begin{equation}
\rho _{e}\left( 0\right) \equiv \left\vert k_{y}\right\vert
\sum_{n=0}^{\infty }c_{n}\left[ \text{sgn}\left( k_{y}\right) \chi
_{xy}+\left( 2n+1\right) \eta _{xx}\right] \ .  \label{re52}
\end{equation}

\begin{figure}[b]
		\includegraphics[width=\columnwidth]{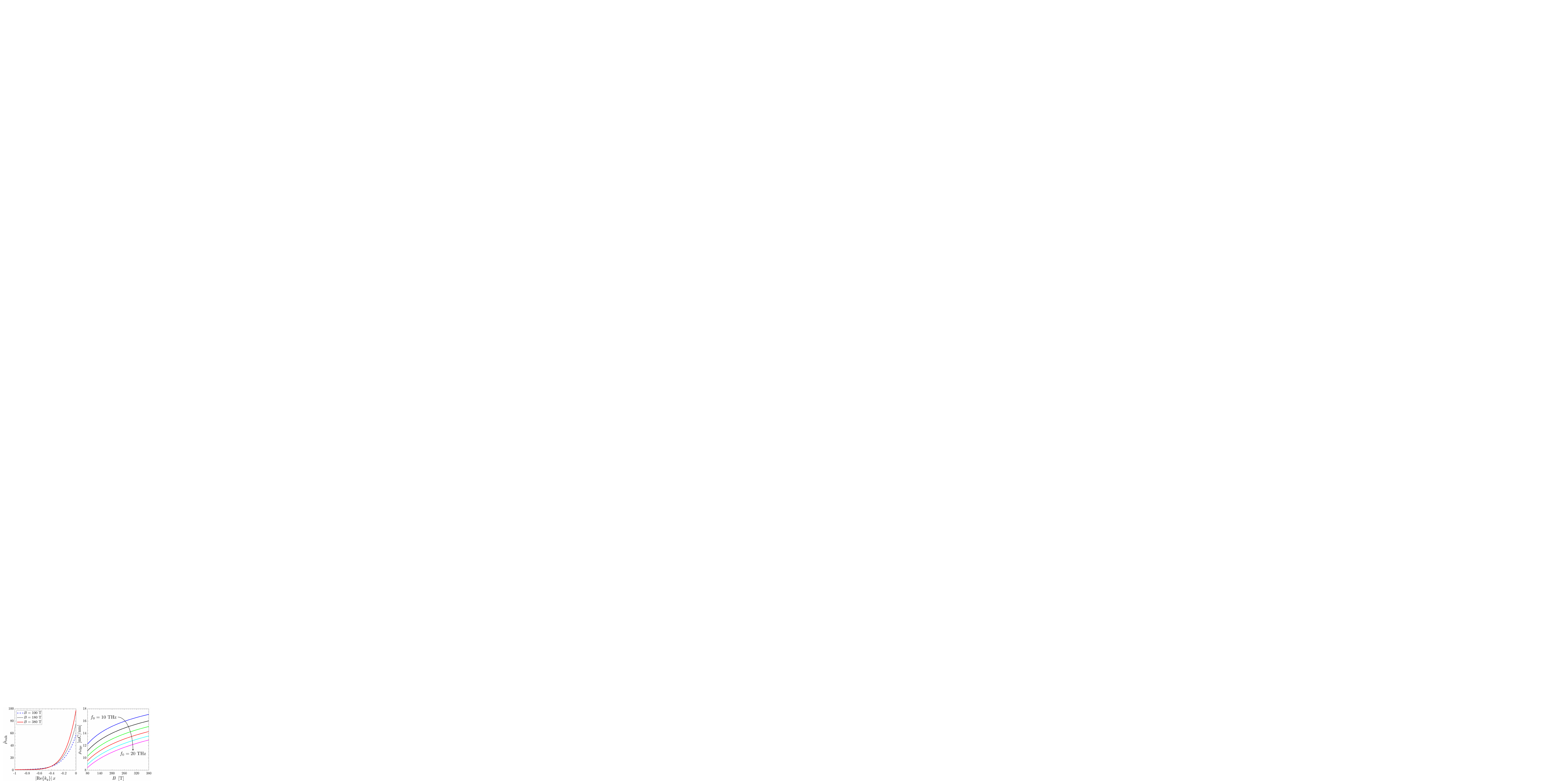}
		\caption{(a) Bulk and (b) edge charge density for graphene in an external magnetic field. $\tilde{\rho}_\textrm{bulk}$ is the bulk charge density normalized by $\rho_\textrm{bulk}$ at $k_x x=-1$, and $\tilde{\rho}_\textrm{edge}$ is the edge charge density normalized by $\rho_\textrm{edge}$ at $B=10$T. $\mu=-0.3$eV, $T=40$K, and $\Gamma=2 \times 10^{12}$/s.} 
		\label{rho}
\end{figure}

Assuming $\sigma _{xx}=\sigma _{yy}$ and $\sigma _{xy}=-\sigma _{yx}$,
\begin{equation}
\rho _{b}\left( y\right) =-\eta _{xx}k_{y}^{2}e^{y}\sum_{n=0}^{\infty }c_{n}%
\frac{d^{2}}{dy^{2}}L_{n+1}\left( -2y\right)  \label{re53}
\end{equation}

Figure~S\ref{rho} shows bulk and edge charge density at several values of external magnetic bias.

\subsection{Approximating the Dispersion Relation}

To a good approximation, the edge dispersion within the first\ TM\ band gap
is obtained by considering only the $n=0$ term in the expansion (\ref{re26}). With the
assumption that $\sigma _{xx}=\sigma _{yy}$ and $\sigma _{xy}=-\sigma _{yx}$%
, we find $J_{00}=0$, leading to%
\begin{equation}
\left\vert k_{y}\right\vert =\left[ 2I_{0}\left( \eta _{xx}\pm \chi
_{xy}\right) \right] ^{-1},  \label{re56}
\end{equation}
where%
\begin{equation}
I_{0}=\left\vert k_{y}\right\vert \int_{-\infty }^{0}dxe^{\left\vert
k_{y}\right\vert x}g\left( x,0\right) ,  \label{re41}
\end{equation}
such that%
\begin{equation}
g\left( x,0\right) =\frac{1}{2\bar{\varepsilon}}\int_{-\infty }^{\infty }%
\frac{dk_{x}}{2\pi }\frac{1}{q}e^{ik_{x}x},  \label{re42}
\end{equation}
which we approximate by expanding $q=\sqrt{k_{x}^{2}+k_{y}^{2}}$ about $%
k_{x}=0$,%
\begin{equation}
\sqrt{k_{x}^{2}+k_{y}^{2}}\simeq \left\vert k_{y}\right\vert +\frac{k_{x}^{2}%
}{2\left\vert k_{y}\right\vert }\ .  \label{re44}
\end{equation}

This leads to the closed form approximate solution of (\ref{re42})%
\begin{equation}
g\left( x,0\right) \simeq g_{0}\left( x,0\right) \equiv \frac{1}{2\bar{%
\varepsilon}\sqrt{2}}e^{-\sqrt{2}\left\vert k_{y}\right\vert \left\vert
x\right\vert },  \label{re43}
\end{equation}
which we use in (\ref{re39}), simplifying the dispersion relation to
\begin{equation}
\left\vert k_{y}\right\vert =\bar{\varepsilon}\frac{1+\sqrt{2}}{\eta
_{xx}\pm \chi _{xy}}\ .  \label{re45}
\end{equation}

We find this result better approximates the exact edge mode dispersion than
that used in previous works \cite{Fetter_2}, \cite{Apell2},%
\begin{equation}
\left\vert k_{y}\right\vert =\bar{\varepsilon}\frac{3\eta _{xx}-\text{sgn}%
\left( k_{y}\right) 2\sqrt{2}\chi _{xy}}{\eta _{xx}^{2}-\chi _{xy}^{2}}\ .
\label{re46}
\end{equation}

\subsection{Material Loss}

\begin{figure}[t]
		\includegraphics[width=0.85\columnwidth]{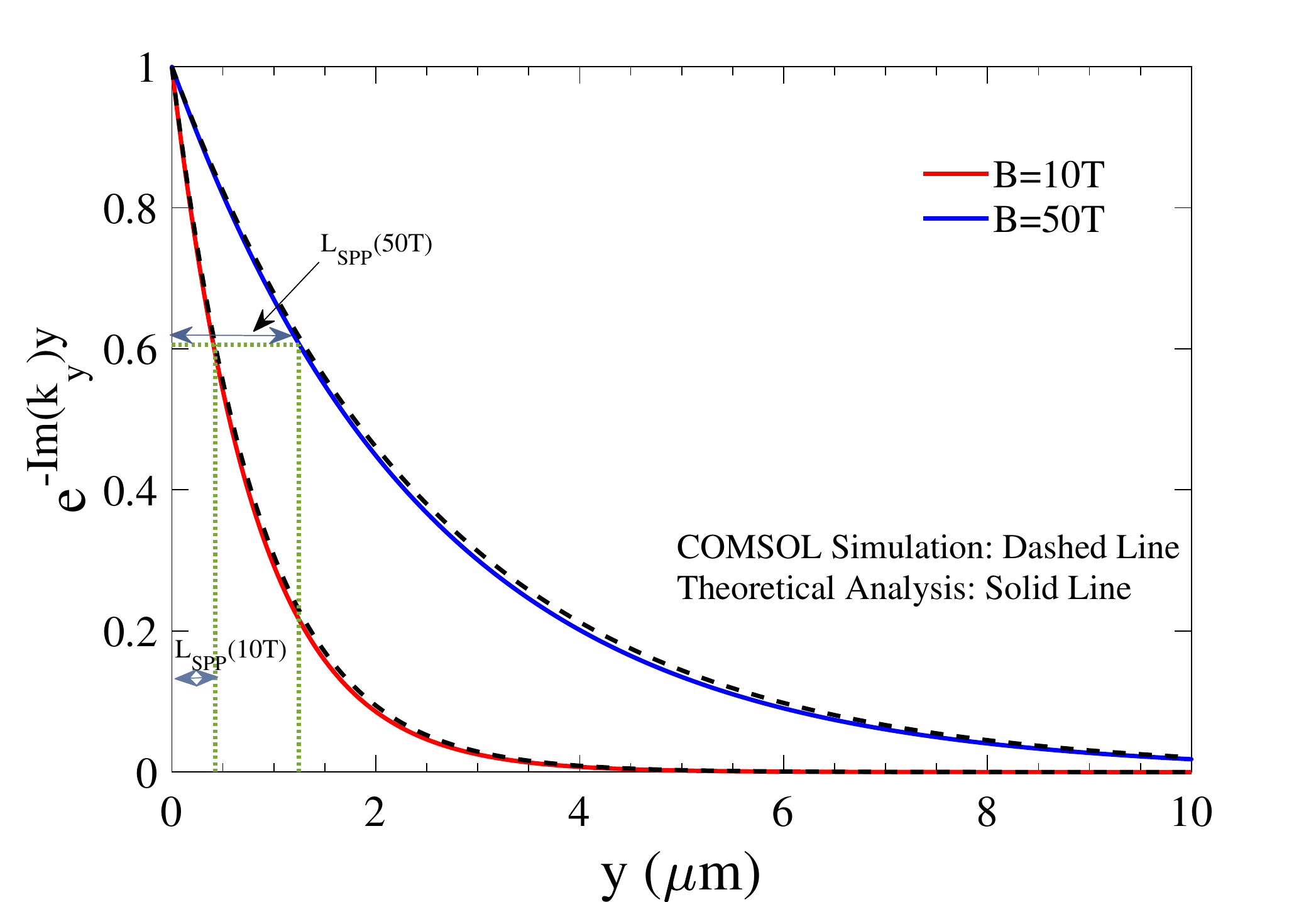}
		\caption{Decay of edge plasmon, comparing perturbative solution for introducing loss (using (\ref{re38})), and full solution of Maxwell's equations (COMSOL) for graphene in an external magnetic field. Frequency is 14 THz and $\mu=0.05$ eV, $T=40$K, $\Gamma=2\times 10^{12}$ rad/s.} 
		\label{DEP2}
\end{figure}

In statics, there is no concept of loss. However, our interest is in the quasi-static regime, such that we can perturb the system slightly by introducing a non-zero scattering rate $\Gamma $ in the
conductivity. Then, we can make the replacement
\begin{equation}
\left\vert k_{y}\right\vert \rightarrow k_{y}=\left\{ 
\begin{array}{cc}
+\textrm{Re}\left( k_{y}\right) +i\textrm{Im}\left( k_{y}\right)  & \textrm{Re}%
\left( k_{y}\right) ,\textrm{Im}\left( k_{y}\right) >0 \\ 
-\textrm{Re}\left( k_{y}\right) -i\textrm{Im}\left( k_{y}\right)  & \textrm{Re}%
\left( k_{y}\right) ,\textrm{Im}\left( k_{y}\right) <0%
\end{array}%
\right. \label{re39}
\end{equation}
which ensures that the wave decays in the case of both forward and
backward propagation. This results in complex-valued wavenumbers for the edge dispersion from both the exact method (\ref{re38}) and from the approximate value (\ref{re45}). As a check, we compared decay rates of the edge SPP
generated using this perturbative approach and the result found via COMSOL. Figure~S\ref{DEP2} shows
good agreement between the two methods for graphene in an external magnetic field.

\begin{figure}[b]
		\includegraphics[width=0.9\columnwidth]{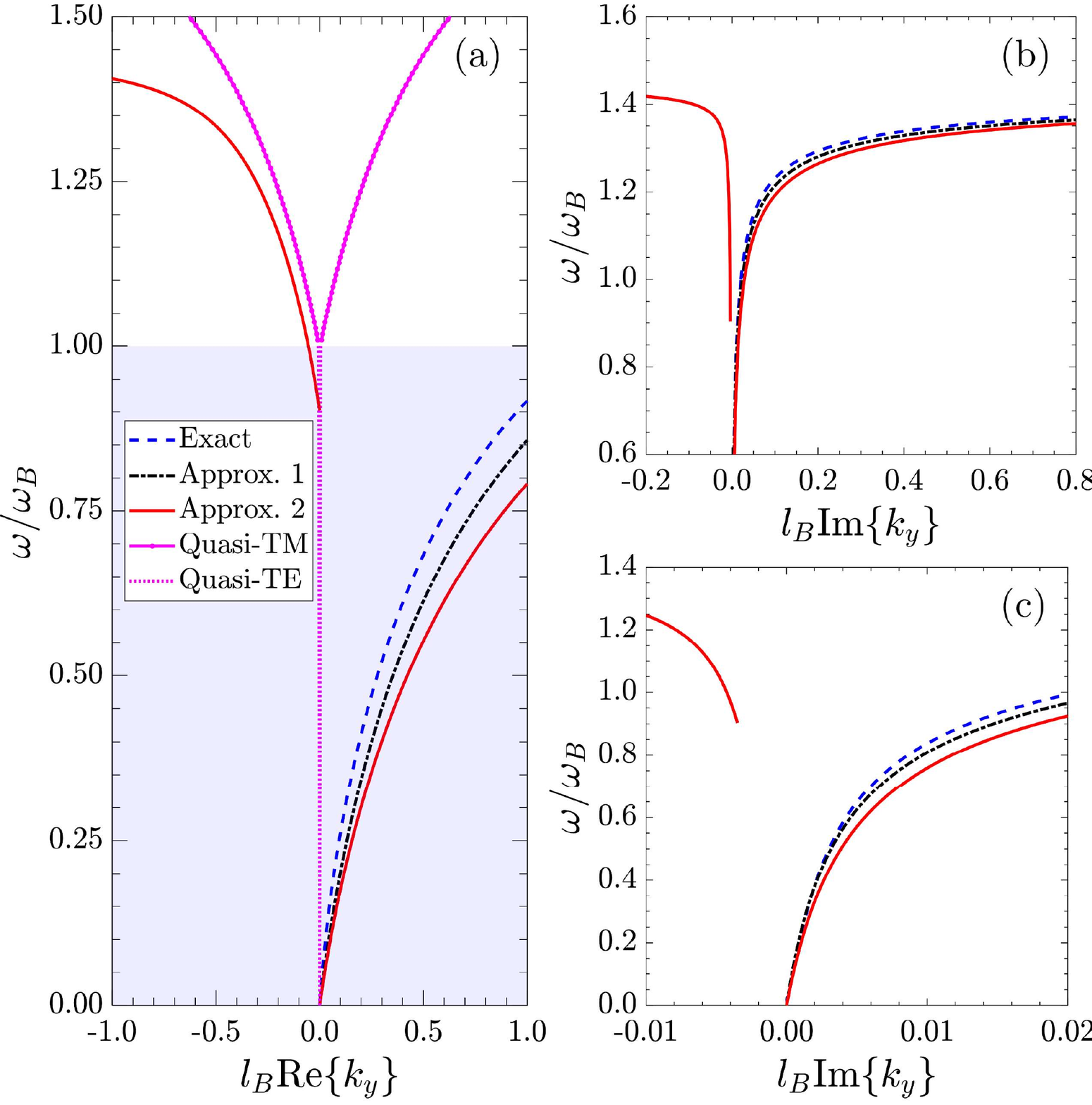}\\
		\caption{Bulk (pink) and edge dispersion of graphene modes for real-part (a) and imaginary part (b)-(c) of wavenumber for graphene in an external magnetic field. The shaded region indicates the bulk band gap, and $\omega_B$ is the frequency of the first Landau level. Approx. 1 is using (\ref{re45}) and Approx. 2 is using (\ref{re46}); $\mu=-0.3$eV, $T=40$K, $\Gamma=2 \times 10^{12}$/s, $B=100$T.  $l_{B}=\sqrt{\hslash /eB}$ is the magnetic length.} 
		\label{Disp1}
\end{figure}

Figure~S\ref{Disp1} shows the bulk and edge dispersion for graphene in an external magnetic bias field. The edge modes were computed using the exact quasi-static analysis (Eq.~(\ref{re38})), and a comparison between the exact and approximate edge dispersion solutions is also shown. Although the results were computed assuming $B=100$T, due to the normalization the dispersion diagrams are essentially independent of $B$ for $\left\vert B \right\vert \gtrapprox 1$ T.  

\section{Edge Mode Model: Charge Density Approach}

The edge dispersion can be determined via an alternative method using the charge density.

\subsection{2D Bulk Mode}

The quasi-static edge mode is obtained using an expansion of the charge density, rather than the potential, in Laguerre polynomials. 
Starting with a 2D conductivity tensor
\begin{equation}
\boldsymbol{\sigma} = \left[\begin{array}{cc}
\sigma_{xx} & \sigma_{xy} \\
\sigma_{yx} & \sigma_{yy}
\end{array}
\right]\end{equation}
with $ \sigma_{xx} = \sigma_{yy} $ and $ \sigma_{yx} = - \sigma_{xy} $, it is assumed that the total electron density can be represented by $ n_0 + n $, where $ n_0 $ is the ground state electron density and $ n $ is the corresponding fluctuation ($ \left\lvert{n}\right\rvert \ll \left\lvert{n_0}\right\rvert $). The electron fluid is confined to the $ z = 0 $ plane. From charge conservation, $\nabla \cdot \boldsymbol{j} = -i e \omega n \nonumber $, and $ \boldsymbol{j} = \boldsymbol{\sigma} \cdot \boldsymbol{E} $,
\begin{equation}\label{beginning}
n = \frac{1}{-ie\omega} \left[  \sigma_{xx} \left( - \frac{\partial^2}{ \partial y^2 }  - \frac{\partial^2}{ \partial x^2 }  \right) \phi - \frac{ \partial \sigma_{xx} }{ \partial x } \frac{ \partial \phi  }{ \partial x } -  \frac{ \partial \sigma_{xy} }{ \partial x } \frac{ \partial \phi  }{ \partial y }    \right]
\end{equation}
where $ \phi $ is the electrostatic potential, related to the electric field by $ \boldsymbol{E} = -\nabla \phi $. 

We first solve for the bulk modes of a laterally-infinite 2D electron fluid. In this case, Poisson's equation is
\begin{equation}
\nabla^2 \phi = \frac{ne}{\epsilon_0} \delta(z).
\end{equation}
We assume plane wave solutions proportional to $ e^{i(\boldsymbol{q} \cdot \boldsymbol{r} - \omega t  )  } $, where $ \boldsymbol{q} $ is the in-plane wavevector, and where the potential amplitude $ \phi(z) $ depends on the distance from the plane as well as on the in-plane wavevector. It is straightforward to show that the solution has the form
\begin{equation}\label{Poisson_solution}
\phi_q(z) = - \frac{  n_q e }{2 \epsilon_0 q } e^{-q|z|} ,
\end{equation}
where $ n_q $ is the corresponding amplitude for the electron density fluctuation. Note that in the quasi-static case the in-plane wavenumber $q$ also serves to govern vertical decay. Assuming no disruption in the conductivity in the $x-y$ plane, $\partial \sigma_{ij} / \partial x =0$, and Eq.~(\ref{beginning}) can be simplified as $ n = \left(\sigma_{xx} / ie\omega\right)  \nabla^2 \phi  $. Replacing $ \phi $ by Eq.~(\ref{Poisson_solution}) yields the equation for the 2D quasi-static bulk SPP modes,
\begin{equation}
\omega \epsilon_0 + \frac{i q \sigma_{xx} }{2 }  =  0 .
\end{equation}

\subsection{2D Edge Mode}

Here we suppose the 2D electron fluid is confined to $z = 0$ and occupies the half-plane on the negative side of the $x$-axis. From Eq.~(\ref{beginning}) we obtain
\begin{align}
i e \omega n = & \, \sigma_{xx} \nabla^2 \phi - \sigma_{xx} \delta(x) \frac{ \partial \phi }{ \partial x  } - \sigma_{xy} \delta(x) \frac{ \partial \phi }{ \partial y } \nonumber \\ 
 = & \, \sigma_{xx} \nabla^2 \phi + j_x \delta(x),
\end{align}
where from $ \boldsymbol{j} = \boldsymbol{\sigma} \cdot \boldsymbol{E} $ we obtain $ j_x = -\sigma_{xx} \partial_x \phi - \sigma_{xy} \partial_y \phi $; the delta functions arise from derivatives of $\sigma_{i,j}=\sigma_{i,j} U(-x)$, where $U(x)$ is the unit step function. The above equation represents a singularity for the current along the $x$-axis at the edge, suggesting nonzero charge accumulation at the edge, $ n = n_b(x) + n_e\delta(x) $ where $ n_b(x) $ is the fluctuation in the bulk and $ n_e $ represents the accumulation at $x = 0 $. Replacing $n$ by $ n_b (x) + n_e \delta(x) $ in the above equation,
\begin{align}
& i e \omega (n_b + n_e \delta(x)) = \sigma_{xx} \nabla^2 \phi - \sigma_{xx} \delta(x) \frac{ \partial \phi }{ \partial_x  } - \sigma_{xy} \delta(x) \frac{ \partial \phi }{ \partial y }. \nonumber
\end{align}
Equating the non-singular terms on the left and right hand sides of the above equation leads to
\begin{equation}\label{ODE}
\left( \frac{ \partial^2 }{  \partial x^2 }   - q^2  \right) \phi = \frac{i e \omega}{\sigma_{xx}} n_b (x),
\end{equation}
where $q=q_y$ here and below. Equating the singular terms leads to 
\begin{equation}\label{ODE_BC}
\left[\left( \frac{ \partial }{ \partial x }  + iq \frac{\sigma_{xy}}{\sigma_{xx}}  \right) \phi\right]_{x = 0^-} = -\frac{ie\omega}{\sigma_{xx}} n_e,
\end{equation}
which can serve as a boundary condition (the left side in the above equation is $j_x(x=0)$). This boundary condition relates the charge accumulation $n_e$ to the normal component of the current at the edge. A Green function approach can be used to solve the above equation \cite{George}. We define a Green function as 
\begin{equation}
\left( \frac{\partial^2}{ \partial x^2 } - q^2 \right) G(x,x') = - \delta(x-x')
\end{equation}
valid for $ x,x' < 0 $ and subject to the following boundary condition
\begin{equation}\label{GF_BC}
\left[\left( \frac{ \partial }{ \partial x }  + iq \frac{\sigma_{xy}}{\sigma_{xx}}  \right)G(x,x')\right]_{x = 0^-} = 0,
\end{equation}
assuming the above homogeneous boundary condition at $ x = 0^- $, a bounded response at $ x \rightarrow - \infty $, and the following jump condition at $ x = x' $,
\begin{equation}
\left(\frac{\partial G(x,x')}{ \partial x }\right)_{x=x'^{+}} - \left(\frac{\partial G(x,x')}{ \partial x }\right)_{x=x'^{-}} = -1.
\end{equation}
The solution for the Green function is
\begin{equation}
G(x,x') =  \frac{\sigma_{xx} - i\sigma_{xy}}{2q(\sigma_{xx} + i\sigma_{xy})} e^{q(x+x')} + \frac{1}{2q} e^{-q|x-x'|},
\end{equation}
and using Green's second theorem the potential can be obtained as
\begin{align}\label{phi_vs_GF}
\phi(x) =& - \frac{i e \omega }{ \sigma_{xx} } \int_{x' = -\infty}^{ 0 } G(x,x')n_b(x')dx' \nonumber \\& +  G(x,0) \left[ \left( \frac{\partial }{\partial x} -  iq\frac{\sigma_{xy}}{\sigma_{xx}}\right) \phi(x) \right]_{x = 0}.
\end{align}

The edge charge accumulation in Eq.~(\ref{ODE_BC}) is in terms of $ \phi(x = 0) $ and $ \partial_x \phi\lvert_{x=0} $. These parameters can be found using the above equation as
\begin{eqnarray}\label{phi0}
\phi(0) & =& \Bigg( -\frac{i e \omega }{ \sigma_{xx} } \int_{x' = -\infty}^{ 0 }
      G(0,x')n_b(x')dx' +  \nonumber \\
     & & G(0,0)\left[  \frac{\partial \phi}{ \partial x } \right]_{0^-} \Bigg)
   \times \left[ 1 - iq\frac{G(0,0)\sigma_{xy}}{\sigma_{xx}} \right]^{-1} 
\end{eqnarray}
\begin{eqnarray}\label{dphi0}
\left[\frac{\partial  \phi(x) }{  \partial x }\right]_{x = 0} & = &
\Bigg(-\frac{i e \omega }{ \sigma_{xx} } \int_{x' = -\infty}^{ 0 }  \frac{\partial
	G(x,x')}{ \partial x }  n_b(x')dx' \nonumber \\
 &&+ \left(\frac{ \partial G(x,0) }{ \partial x } \right)_{x = 0}  i q \frac{\sigma_{xy}}{
\sigma_{xx} } \phi(x = 0) \Bigg) \nonumber \\
 && \times \left[1 - \left(\frac{ \partial G(x,0) }{ \partial x }\right)_{x = 0} \right]^{-1}
\end{eqnarray}
and simultaneously solving equation \ref{phi0} and \ref{dphi0} for $ \phi(x = 0) $ and $ \partial_x \phi\lvert_{x=0} $ gives

\begin{widetext}
\begin{equation}
\left[\frac{\partial \phi(x)}{ \partial x }\right]_{x = 0} = \frac{1}{ \alpha \gamma -XG(0,0) } \left( \frac{-ie\omega}{ \sigma_{xx} }  \right) \int_{x'=-\infty}^{0 } \left[\alpha \left(\frac{\partial G(x,x')}{ \partial x }\right)_{x = 0} + XG(0,x')\right] n_b(x')dx'
\end{equation}
\begin{equation}
\phi(x = 0) = \frac{-ie\omega}{ \sigma_{xx} } \int_{x'=-\infty}^{0} \left[ \left(\frac{1}{\alpha} + \frac{ G(0,0) X }{ \alpha \left( \alpha \gamma - XG(0,0) \right) }\right) G(0,x') + \frac{G(0,0)X}{ \alpha \gamma - XG(0,0) } \left(\frac{ \partial G(x,x') }{\partial x}\right)_{x = 0} \right]  n_b(x')dx'
\end{equation}
\end{widetext}
where 
\begin{align}
& \alpha = 1 - G(0,0) iq \frac{\sigma_{xy}}{ \sigma_{xx} }, ~~\gamma = 1 - \left(\frac{ \partial G(x,0) }{ \partial x }\right)_{x = 0} \nonumber \\ &
X = \left(\frac{ \partial G(x,0) }{ \partial x }\right)_{x = 0}   i q \frac{\sigma_{xy}}{ \sigma_{xx} }.
\end{align}
Inserting $ \phi(x = 0) $ and $ \partial_x \phi\lvert_{x = 0} $ in (\ref{ODE_BC}) gives $ n_e $ in terms of the Green function as
\begin{widetext}
\begin{equation}\label{ne}
n_e = \int_{x'=-\infty}^{0} \left[\left(A \alpha  + Biq\frac{\sigma_{xy}}{\sigma_{xx}}\right) \frac{a-1}{2} e^{qx'} + \left(A X  + C iq\frac{\sigma_{xy}}{\sigma_{xx}}\right) \frac{a+1}{2q} e^{qx'}\right] n_b(x')dx',
\end{equation}
\end{widetext}
where 
\begin{align}
& a =  \frac{\sigma_{xx} - i\sigma_{xy}}{\sigma_{xx} + i\sigma_{xy}}, ~~
A = \frac{1}{ \alpha \gamma - XG(0,0) } \nonumber \\ &
B = \frac{XG(0,0)}{\alpha \gamma - XG(0,0) }, ~~
C = \frac{1}{\alpha} + \frac{ G(0,0) X }{ \alpha \left( \alpha \gamma - XG(0,0) \right) }.
\end{align}
It is shown below that the potential satisfies the integro-differential equation

\begin{align}\label{IE}
\phi(x,z=0) = & - \frac{e}{ \epsilon_0 }  \int_{x' = -\infty}^{  0  } dx' L(x, x')n_b(x') \nonumber \\& - \frac{e}{ \epsilon_0 } L(x,0) n_e
\end{align}
where 
\begin{equation}
L(x,x') = \int_{k = -\infty}^{ + \infty}  \frac{dk}{2 \pi} \frac{ e^{ik(x - x')}  }{2 \sqrt{k^2 + q^2} }.
\end{equation}
Alternatively, one may use the approximate expression $L_0(x,x') = 2^{-3/2} e^{ -\sqrt{2}q|x-x'| }$\cite{Fetter_1,Fetter_2}. Replacing the potential by the expression of potential in Eq.~(\ref{phi_vs_GF}) gives
\begin{align}\label{n_b_int_eq}
& \frac{i  \omega \epsilon_0}{ \sigma_{xx} } \int_{x' = -\infty}^{ 0 } G(x,x')n_b(x')dx' -   \int_{x' = -\infty}^{  0  } dx' L(x, x')n_b(x') \nonumber \\&  + \left(G(x,0) \frac{i \omega \epsilon_0}{ \sigma_{xx} }  - L(x,0)\right) n_e = 0,
\end{align}
where $n_e$ (charge accumulation at the edge) is given by Eq.~(\ref{ne}). We expand the bulk charge fluctuation $n_b $ in terms of Laguerre polynomials \cite{Quinn,Apell}
\begin{equation}\label{Laguerre}
n_b = \sum_{j = 0}^{ \infty } b_j e^{qx} l_j(-2qx)
\end{equation}
subject to the orthogonality relation
\begin{equation}
\int_{x  = -\infty}^{ 0 }  (e^{qx} l_i(-2qx)) \times (e^{qx} l_j(-2qx)) dx = \frac{\delta_{ij}}{2}.
\end{equation}
The integral equation in (\ref{n_b_int_eq}) can be written in a matrix form in terms of unknown constants $ b_j $ and $ n_e $,

\begin{equation}\label{matrix_equation}
\sum_{j=0}^{ \infty } \left[ \frac{i\omega \epsilon_0}{ \sigma_{xx} } G_{ij} -  L_{ij}  + Ne_{ij}  \right]b_j = 0
\end{equation}
where 
\begin{widetext}
\begin{align}
& G_{ij} = \int_{x = -\infty}^{0} \int_{x' = -\infty}^{0} dx dx' e^{qx}  e^{qx'}   l_i(-2qx) G(x,x')  l_j(-2qx') \nonumber \\ &
L_{ij} = \int_{x = -\infty}^{0} \int_{x' = -\infty}^{0} dx dx' e^{qx}  e^{qx'}   l_i(-2qx) L(x,x')  l_j(-2qx')\nonumber \\ &
Ne_{ij} = \int_{x = -\infty}^{0} \int_{x' = -\infty}^{0} dx dx' e^{qx}  e^{qx'}   l_i(-2qx) \Lambda \left(    \frac{i \omega \epsilon_0}{ \sigma_{xx} } \frac{a + 1}{2q} e^{q(x+x')}  -  L(x,0) e^{qx'} \right)  l_j(-2qx') \nonumber \\&
\Lambda =   \frac{a - 1}{2}  \left( A \alpha + Biq\frac{\sigma_{xy}}{\sigma_{xx}}  \right) +  \frac{a + 1}{2q}  \left( A X + C iq\frac{\sigma_{xy}}{\sigma_{xx}}  \right)   .
\end{align}
\end{widetext}
Finally, forcing the determinant of (\ref{matrix_equation}) to be zero leads to the edge mode dispersion equation.

\textbf{Derivation of IE}:
The Poisson's equation considering edge mode propagating along the $y$-axis as $ e^{iq_yy}, ~ q_y = q $, is
\begin{equation}
\left(\frac{\partial^2}{\partial x^2} + \frac{\partial^2}{\partial z^2} - q^2\right) \phi =  \frac{n e}{\epsilon_0} \delta(z) \Theta (-x),
\end{equation}
and taking a Fourier transform along the $x$-axis leads to
\begin{equation}
\left[\frac{\partial^2}{\partial z^2} - \left(k^2 + q^2\right) \right] \phi =  \frac{e}{\epsilon_0} \delta(z) n(k),
\end{equation}
where $ n(k) $ is the Fourier transform of $ n \Theta(-x) $. The potential takes the form
\begin{equation}
\phi(k,z) = -  \frac{e}{2 \epsilon_0} \int_{x'= -\infty}^{0} n(x') e^{-ikx'} dx' \frac{1 }{ \sqrt{ k^2 + q^2 } } e^{- \sqrt{ k^2 + q^2 } |z| } .
\end{equation}
By defining the function kernel $ L(x) $ as
\begin{equation}
L(x) = \int_{k = -\infty}^{ + \infty}  \frac{dk}{2 \pi} \frac{ e^{ikx}  }{2 \sqrt{k^2 + q^2} },
\end{equation}
then the potential function on the surface of the 2D electron fluid becomes
\begin{equation}\label{GF_response}
\phi(x,z = 0) = - \frac{e}{\epsilon_0}    \int_{x' = -\infty}^{  0  } dx' L(x-x')n(x').
\end{equation}
Replacing $n(x)$ by $ n_b(x) + n_e \delta(x) $ gives
\begin{align}
\phi(x,z = 0)  = & - \frac{e}{\epsilon_0}     \int_{x' = -\infty}^{  0  } dx' L(x,x') \left[   n_b(x') + n_e \delta(x') \right] \nonumber \\&
- \frac{e}{\epsilon_0}     \int_{x' = -\infty}^{  0  } dx' L(x,x')   n_b(x') - \frac{e}{\epsilon_0} L(x,0)n_e,
\end{align}
and substituting $ n_e $ from the boundary condition equation leads to (\ref{IE}).